\documentclass[a4,12pt]{article}
\usepackage{graphicx}
\usepackage{subfigure}
\usepackage{amsmath}
\usepackage{amsfonts}
\usepackage{amssymb}

\usepackage{vmargin}
\setmarginsrb{2cm}{2cm}{2cm}{2cm}{0.7cm}{0.5cm}{0.7cm}{0.5cm}

\begin{document}

\title{{\bf Aging Dynamics of a Fractal Model Gel}}

\author{%
  M-A. Suarez$^{1,2}$, N. Kern$^1$, E. Pitard$^{1,3}$, and W. Kob$^1$\\
  \begin{minipage}[t]{\textwidth}
    \null$^1$\textit{%
      Laboratoire des Collo\"ides, Verres et Nanomat\'eriaux, 
      UMR 5587,
      Universit\'e Montpellier 2, F-34095 Montpellier, 
      France}\\
    \null$^2$\textit{%
      Facultad Experimental de Ciencias y Tecnolog\'{\i}a - FACYT, 
      Departamento de F\'{\i}sica, Universidad de Carabobo, 
      Valencia 2001, 
      Venezuela}\\
    \null$^3$\textit{
      CNRS, UMR 5587, 
      F-34095 Montpellier, 
      France}
    \end{minipage}
}

\date{}

\maketitle

\abstract{
Using molecular dynamics computer simulations we investigate the aging
dynamics of a gel. We start from a fractal structure generated by the
DLCA-DEF algorithm, onto which we then impose an interaction potential
consisiting of a short-range attraction as well as a long-range
repulsion. After relaxing the system at $T=0$, we let it evolve at
a fixed finite temperature. Depending on the temperature $T$ we find
different scenarios for the aging behavior.  For $T\gtrsim 0.2$ the
fractal structure is unstable and breaks up into small clusters which
relax to equilibrium. For $T\lesssim 0.2$ the structure is stable and
the dynamics slows down with increasing waiting time. At intermediate
and low $T$ the mean squared displacement scales as $t^{2/3}$ and we
discuss several mechanisms for this anomalous time dependence. For
intermediate $T$, the self-intermediate scattering function is given
by a compressed exponential at small wave-vectors and by a stretched
exponential at large wave-vectors.  In contrast, for low $T$ it is a
stretched exponential for all wave-vectors. This behavior can be traced
back to a subtle interplay between elastic rearrangements, fluctuations
of chain-like filaments, and heterogeneity.
}

\section{Introduction}

Gels are mechanically stable structures which have a low volume
fraction. They are usually made of colloids or of macromolecules, such
as assemblies of proteins, clays or of a colloid-polymer mixture. Their
structure can be characterized as an open but percolating network of
particles having a fractal dimension~\cite{weitz_fractal}. Due to their
importance in technology and fundamental science, gels have been studied
extensively by means of experiments and theory as well as numerical
simulations \cite{reviewPuertas-Fuchs}.

Various models have been proposed to describe gels on the microscopic
level.  Predominantly, a short-range attractive potential between
particles is used, often complemented by a long-range repulsion
\cite{delgado, sciortino}. Gels have also been obtained from so-called
``patchy'' colloids, in which the particles do not interact through a
spherically symmetric potential, but rather carry ``active spots'' in
which a strong attraction is localised~\cite{active spots,delgadokob}.
Alternatively, a gel can be produced by requiring that the maximum
number of neighbors of any given particle remain below a given
threshold~\cite{maxneighbours} or by introducing many-body interactions
\cite{delgadokob}. Note that all these interactions are very different
from the one found in dense glass-forming systems (atomic or molecular)
for which the relevant potential is dominated by a hard sphere-like
interaction~\cite{pusey,vanmegen}.

Furthermore, the question about the thermodynamic stability of  gels
has been extensively studied recently. In particular it is still a
matter of debate in which cases a gel results from an arrested phase
separation process, and in which cases it may arise as a truly stable
thermodynamical phase \cite{delgadokob,zaccarellireview,nature_lu}.

In practice, gels are often out-of-equilibrium systems and, therefore,
undergo aging. Some of them are entirely impossible to synthesize in
equilibrium.  In contrast to the structure and relaxation dynamics of
gels, which have been extensively studied, much less is known on their
aging dynamics \cite{faraday,cipelletti-2005-17}, with the exception
of aging in dense glass-formers, which benefit from previous studies
in spin-glasses and disordered systems \cite{leticia}. It is remarkable
that in these systems quantities like the dynamic response do show strong
aging dynamics whereas the structure hardly changes during aging since
it is only weakly dependent on $T$.

The goal of the present work is to investigate the aging dynamics of a
fractal gel. Such out-of-equilibrium studies of gels are rare, although
a few exist, using an approach which is slightly different from ours
\cite{lemans}.  Here, we investigate how the structure and the dynamics
change with the age of the system and show that, depending on temperature,
different aging regimes can be distinguished. In Sec.~2 we give the
details of the model we have used and of the simulations. In Sec.~3
we will present the results for the structural quantities, and in the
subsequent section~4 we discuss the relaxation and aging dynamics. Finally
we conclude with a summary of our results and discuss some open questions.

\section{Model and Simulation Methods}

Experiments show that colloidal gels occur at low volume fraction
\cite{reviewPuertas-Fuchs,cipelletti-2005-17,PhysRevLett.84.2275} and show
a fractal-like open network structure which is usually preserved over
a timescale of a typical experiment. Our approach consists therefore of
three steps, the details of which will be discussed below in more detail:
First, we construct a fractal initial configuration, using a purely
kinetic particle diffusion algorithm. Second, we switch on an interaction
potential between the particles and allow the initial structure to relax
locally in order to adapt to the interactions.  Finally, we follow the
dynamics of the system via constant temperature Molecular Dynamics (MD)
simulations in order to investigate the structural and the dynamical
properties of the system as a function of the waiting time.

\paragraph{Initial configuration:}
We construct the initial fractal configuration for our MD simulations
using the so-called ``diffusion limited cluster aggregation''
(DLCA) algorithm \cite{PhysRevLett.51.1119, PhysRevLett.51.1123,
PhysRevB.50.6006}.  This procedure involves letting each particle diffuse
freely until it touches another particle, i.e. until the distance to
the other particle is equal to a certain fixed length, $R_{min}$. The
touching particles then form a permanent and rigid bond. The resulting
cluster diffuses as such and can subsequently  collide and bind to
another particle or particle cluster.  This algorithm is known to lead
to fractal structures of rather low volume fraction. In practice, we
have used a method which combines the algorithm just described with the
so-called ``Dangling Bond Deflection algorithm'' (DLCA-DEF algorithm)
\cite{PhysRevE.65.041403}, which includes rotational diffusion of
clusters around particles chosen at random, thereby allowing the
formation of extended loops in the network structure. The configurations
obtained in this way not only retain the fractal structure created
from DLCA, but also have more realistic structural and mechanical
properties such as the dependence of the elastic modulus on density
\cite{PhysRevE.65.041403,JourNon-CrystSol_Ma_Jullien}.  Specifically,
we produce our configurations with $N=2388$ particles, at a particle
density $\rho=0.019$, and a maximum coordination number $c_{max} = 4$.
The resulting configurations are only retained if they percolate with
respect to a periodic simulation box in all three spatial directions,
which is the case in roughly $25\, \%$ of the cases.

\paragraph{Interaction potential and initial relaxation:}

The cluster obtained by the DLCA-DEF algorithm is a purely geometrical
object: The particles are permanently bound and do not evolve under
the effect of an interaction potential. In order to be able to study
the {\it dynamical} properties of such a cluster it is therefore
necessary to introduce such an interaction, but without jeopardizing
the fractal network. In the past it has been recognized that open
network structures resembling those found in real gels can be obtained
by using an interaction potential involving a pure two-body term with
an attractive well as well as a repulsive barrier at somewhat larger
distances~\cite{delgado,sciortino}, even though this possibility is
not the only one~\cite{active spots, maxneighbours,delgadokob}. The
attractive well is needed in order to give cohesion to the structure. The
barrier, on the other hand, is necessary in order to energetically
penalize certain inter-particle distances, thereby preventing the open
structure of the gel phase from becoming thermodynamically unstable
with respect to a collapse via creation of a dense phase and subsequent
phase separation  into gas and liquid. At the same time the presence
of the barrier  also contributes to preventing the disintegration of
the cluster~\cite{sciortino}.  The interaction potential we use thus
has a repulsion term of a soft sphere and two additional terms for the
attractive well and the repulsive barrier:

\begin{equation}
\label{eq1}  
  V(r) = \epsilon \left[ \left(\frac{\sigma}{r}\right)^{12} 
    - A_{0} \, g(r; R_0, l_0, f_0)
    + A_{1} \, g(r; R_1, l_1, f_1)
  \right] \quad .
\end{equation}

\begin{figure}
\centering
\includegraphics[scale=0.8]{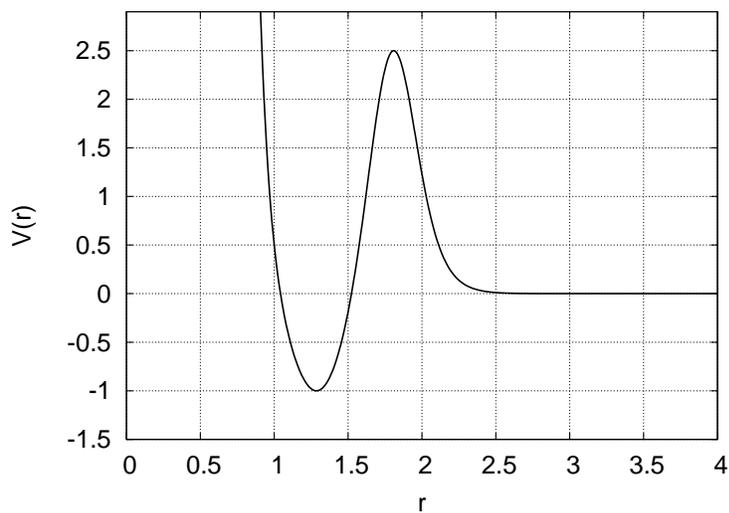}
\caption{
The potential used in the molecular dynamics consists of a repulsive core
of a soft sphere, an attractive potential well and a repulsive barrier
which stabilizes the structure. See Eqs.~(\ref{eq1}) and (\ref{eq2})
for details.
}
\label{Fig_Potential}
\end{figure}

\noindent
Here, the function

\begin{equation} 
\label{eq2}
  g(r; R,l,f ) 
  = 
  \tanh(\frac{r-R}{l}+f)-\tanh(\frac{r-R}{l}-f)
\end{equation}

\noindent
is used twice to create the peak and the well, and where the parameters
$R_{0,1}$, $l_{0,1}$, and $f_{0,1}$ allow to adjust their position,
width and shape.  For the present work we used the following set of
parameters: $\sigma=1.0$ (i.e. $\sigma$ is our unit of length), $R_0 =
1.3$, $R_1 = 1.8$, $l_0=0.285$, $l_1=0.2$, $f_0=f_1=0.5$, $A_0=1.22$,
$A_1 = 2.87$. These parameters produce a potential with a well of depth
$\epsilon$ (and in the following we will set $\epsilon=1$, i.e. use
$\epsilon$ as our energy unit) which is located at $R_{min} \approx
1.2$9, and a barrier with height of approximately $1.5$ and which
is located at $R_{max} = 1.81$. The full potential is shown in Fig.
\ref{Fig_Potential}. Note that $R_{min}$ is chosen so as to coincide
with the distance separating two particles in contact, as defined in
the DLCA process.

\begin{figure}
\centering
\includegraphics[scale=0.5,angle=0]{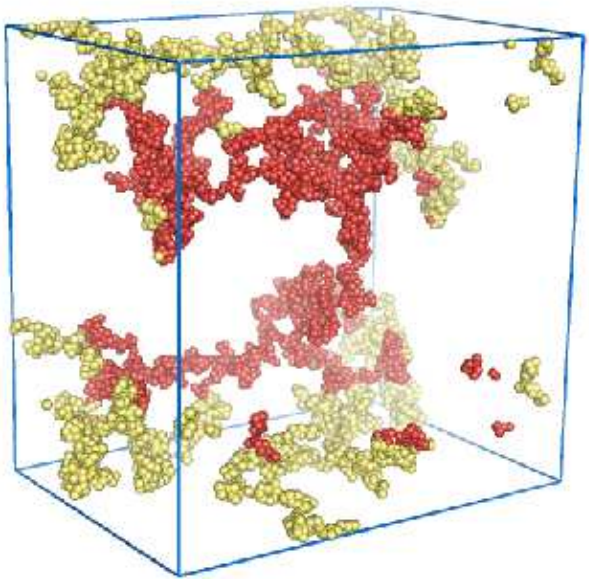}
\caption{
Example of an initial structure used for the simulations.  A fractal
cluster is grown using particle diffusion kinetics and then allowed to
relax to a local minimum under the effect of the particle interactions. We
use periodic boundary conditions and only clusters percolating in all
three spatial directions are retained. The main cluster is shown in dark
gray (red online), whereas slices of periodic copies are reproduced in light gray (yellow online)
in
order to allow to see the percolating structure.
}
\label{Fig_cluster2}
\end{figure}

Although the interaction potential given by Eq.~(\ref{eq1}) is potentially
compatible with gel-like structures, it is evident that the cluster
produced by the DLCA-DEF construction will in general be far from an
energetic optimum since it does not take into account the particle
interaction and in particular we must expect to find particles at,
or close to, the unfavourable distance $R_{max}$ corresponding to the
potential barrier.  Using the DLCA-DEF cluster directly as an initial
configuration for a MD run at constant energy with the interaction
potential given by Eq.~(\ref{eq1}) would therefore very likely lead to
a quick disintegration of the cluster.

In order to avoid this problem we first allowed the DLCA-DEF cluster
to adapt to the interaction between particles by letting it relax
toward a (local) energy minimum. In practice this was done by running
a MD simulation at zero temperature (using a time step of $h=0.001$),
thereby allowing unfavorably placed particles to adapt their positions
before starting the actual simulations. We found that in general the
overall fractal, open structure of the cluster remains unaffected by this
process (details of structural changes at this stage will be discussed
below), and only rarely the relaxation led to the breaking of a chain,
in which case the resulting structure was discarded. We also discarded
the configuration in the case where this adaptation led to a loss of
percolation in any spatial direction.  In Fig.~\ref{Fig_cluster2}
we show a typical resulting cluster used as initial configuration for
the MD simulations at constant temperature.

\paragraph{Molecular Dynamics}
Using the relaxed clusters as initial configurations, we carried out
molecular dynamics simulations to study their structural evolution at
various temperatures ($T=0.05$, $0.10$, $0.15$, $0.20$, $0.25$, $0.30$,
$0.35$, $0.40$, $0.45$, $0.50$, $0.55$ and $0.65$). Furthermore we
investigated dynamic quantities. Since at the start of the simulation
the system is out of equilibrium one has to use two-times quantities in
order to characterize the relaxation dynamics~\cite{leticia}, which depend
on both the waiting time $t_w$ (also called the ``age of the system'',
which is the time which has elapsed between the initial zero-temperature
adaptation of the cluster and the beginning of the dynamical measurement),
as well as the time lag $\tau$ since the beginning of the dynamical
measurement.

As already mentioned,  we use $\sigma$ and $\epsilon$ as units of
length and energy, respectively. Temperature is thus measured in units
of $\epsilon$, whereas time is measured in units of $\sqrt{\sigma^2
m/\epsilon}$, with $m$ being the mass of a particle.

We use the standard  velocity-Verlet algorithm~\cite{Allen_Tildesley}
to propagate the system in phase space.  Since it is out of equilibrium,
coupling it to a heat bath is required in order to prevent it from heating
up. We use a variant of Andersen's thermostat \cite{Allen_Tildesley}
by randomizing periodically all velocities according to the appropriate
Maxwell-Boltzmann distribution (in practice every $\Delta_{therm}=0.25$
time units). Thus this type of dynamics is Newtonian for short times,
but resembles Brownian dynamics for times longer than $\Delta_{therm}$.
For computational efficiency the interaction potential is truncated
and shifted at a cutoff radius of $R_c=2.58$.  All simulations were
carried out with a time step of $h=0.005$, except for intervals of
duration $\Delta_{therm}$ after selected waiting times, where a  time
step of $h=0.001$ was used in order to improve the resolution for the
mean square displacement (see below).

In order to improve the statistics of the results, we have furthermore
averaged over ten runs with independent initial configurations.

\section{Structure}

We start the discussion of the properties of the gel by considering its
structure and its evolution as a function of the waiting time $t_w$.

\paragraph{Radial distribution function}
We first of all analyze the distribution of inter-particle
distances, through the radial distribution function $g(r)$, defined
as~\cite{hansenmacdonald86}

\begin{equation}
g(r) = \frac{1}{4\pi r^2 \rho N}
       \sum_k \sum_{j\neq k} \langle \delta(r-|\vec{r}_j-\vec{r}_k|)\rangle \quad .
\label{eq3}
\end{equation}

\begin{figure}
\centering
\includegraphics[scale=0.08, angle=0]{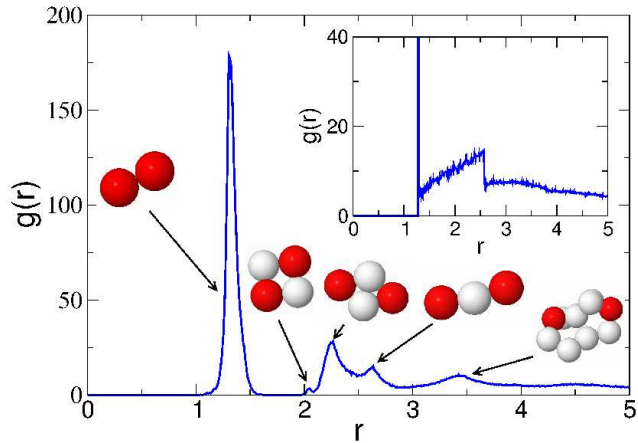}
\caption{
The radial distribution function, $g(r)$, corresponding to the fractal
clusters after initial adaptation to the interaction potential. The main
peaks are labeled by the corresponding local particle arrangements, the
distance $r$ referring to the separation between particles displayed in
dark gray. For comparison, the inset shows $g(r)$ for the configurations
obtained from the DLCA-DEF algorithm, i.e. before adaptation to the
potential.
}
\label{Fig_Structure_Initial_RDF}
\end{figure}

\noindent
The inset in Fig.~\ref{Fig_Structure_Initial_RDF}, shows the radial
distribution function of the original DLCA-DEF structure, i.e. before
the zero-temperature adaptation to the interaction potential given
by Eq.~(\ref{eq1}). We see that the first nearest neighbor peak at
$R_{min}\approx 1.29$ is very high and narrow, as is  to be expected
in view of the DLCA-DEF algorithm used to construct the structure.
Furthermore we notice that $g(r)$ is non-zero for all distances larger
than $R_{min}$ and that there are distances at which singularities
arise, e.g. around $r=2.5$. These features are also directly related
to the algorithm, and they highlight that the DLCA-DEF structure is
very different from that of simple liquids such as hard spheres or
Lennard-Jones particles.

In the main panel of the figure we show the radial distribution
function at $t_w=0$, i.e. immediately after the DLCA-DEF structure has
been allowed to relax to the nearest minimum of the potential given by
Eq.~(\ref{eq1}). At this stage the structure has changed significantly and
several peaks are present, which we have labelled by the local particle
configurations responsible for these peaks.  Due to the low dimensionality
of the structure, this correspondence can be established up to relatively
large distances, unlike what is the case in dense simple liquids. We may
furthermore deduce from this figure that particles have been expelled
from the inter-particle distance corresponding to the position of the
repulsive barrier ($R_{max}\approx 1.8$), as is witnessed through the
drastical suppression of $g(r)$ around these values. In this sense, too,
the local structure of our model gel is therefore very different from
a simple liquid. In the following the presence of this gap in $g(r)$
will allow for a straightforward definition of first nearest neighbor
pairs in the following.

Finally we point out that the location of the main peak is at $r\approx
1.33$, i.e. at a distance slightly larger than the location of the
minimum in $V(r)$ (which occurs at $R_{min}\approx 1.29$), implying that
the system is under weak tension, i.e. that the pressure is negative.

\begin{figure}
\centering
\includegraphics [scale=0.8, angle=0]{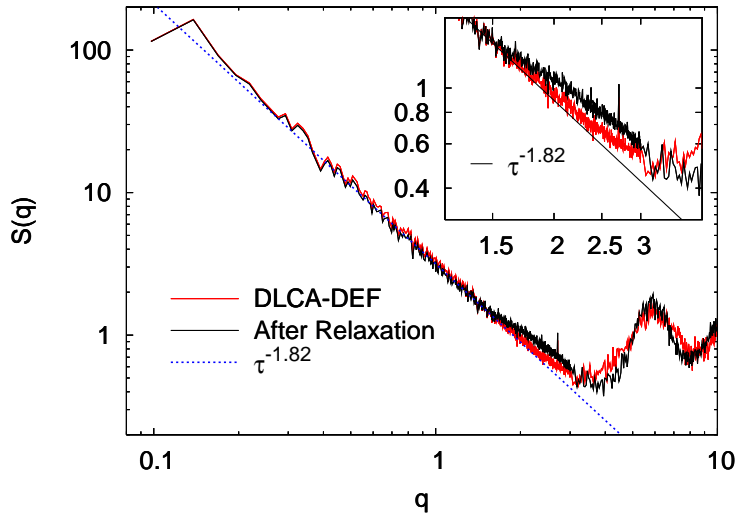}
\caption{
Static structure factor as function of the wave-vector. The thin and bold
lines are for the DLCA-DEF and the relaxed structures, respectively.
The power-law fit yields the fractal dimension.  The inset shows an
enlargement at intermediate $q$ in order to illustrate the change of
the structure on the scale of the filament thickness, i.e. several particle
diameters.
}
\label{Fig_Structure_Initial_SSF}
\end{figure}

\paragraph{Static structure factor}
Complementary information on the structure can be obtained from the
static structure factor $S(q)$ which is given by~\cite{hansenmacdonald86}

\begin{equation}
S(q)= \frac{1}{N}\sum_j\sum_{k} \exp[i \vec{q} \cdot (\vec{r}_j - \vec{r}_k)] \quad ,
\label{eq4}
\end{equation}

\noindent
where $q=|\vec{q}|$ is the modulus of the wave-vector. The $q-$dependence
of $S(q)$ is presented in Fig.~\ref{Fig_Structure_Initial_SSF} where we
compare the static structure factors for the DLCA-DEF cluster before the
zero-temperature relaxation with the one after the relaxation.  We see
that the small wave-vector regimes are identical within the noise of the
data, and thus we conclude that the relaxation process has not changed
the large scale structure of the cluster.  On these length scales the
$q-$dependence is described well by a power-law, $S(q) \sim q^{-d_f}$,
with an exponent $d_f\approx 1.8$. This confirms that the system is
fractal, as is the case for experimentally observed colloidal gels
\cite{PhysRevLett.84.2275,Pine2000}. Effects of the zero-temperature
relaxation  are visible at intermediate and large wave-vectors ($q \geq
1$) showing that rearrangements arise on the scale of a few particles,
in agreement with the conclusions from $g(r)$.  However, the differences
in $S(q)$ are much less pronounced than the ones found in $g(r)$ (see
Fig.~\ref{Fig_Structure_Initial_RDF}), which shows that in this case the
radial distribution function is more sensitive to the modifications than
the static structure factor.  From Figs.~\ref{Fig_Structure_Initial_RDF}
and \ref{Fig_Structure_Initial_SSF} we can conclude that the structure
of the  DLCA-DEF clusters as well as the relaxed clusters form an
open and percolating network structure with a fractal dimension,
which is compatible with the structure found in real colloidal gels
\cite{PhysRevLett.84.2275}.

\begin{figure}
\centering
\includegraphics[scale=0.8, angle=0]{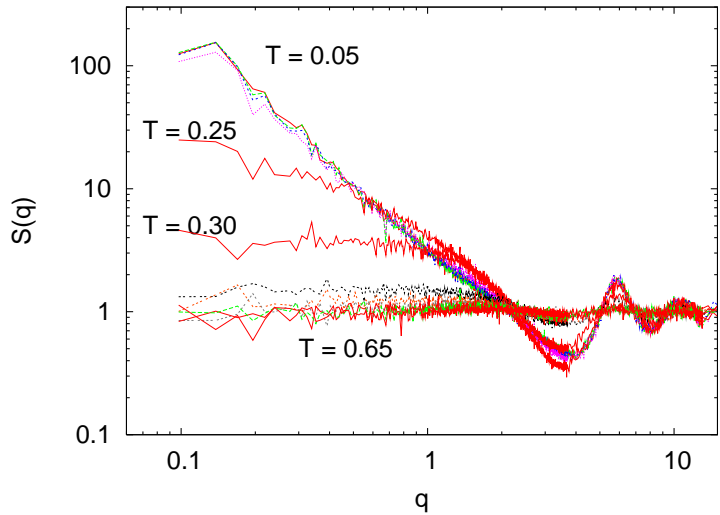}
\caption{
Static structure factor $S(q)$ after the waiting time time $t_w \approx
75000$ for all temperatures considered ($T=0.05, ... , 0.65$, with an
increment of $0.05$). For moderate temperatures the fractal structure is
lost on large scales, but remains intact over intermediate distances. For
higher temperatures, all structure is lost as the initial cluster decomposes
into a gas of particles and small clusters.
}
\label{Fig_Structure_SSF_Evolution}
\end{figure}

The zero-temperature adaptation to the interaction potential can,
however, only be expected to produce a metastable structure, which will
evolve with time. In the following we therefore discuss the waiting
time dependence of $S(q)$.  In Fig.~\ref{Fig_Structure_SSF_Evolution}
we show $S(q)$ for a waiting time $t_w\approx 75000$. The different
curves correspond to different temperatures and we see that $S(q)$ for
low $T$ is very different from the one at high $T$. The structure for
the lowest temperatures has not changed much from the relaxed initial
configuration, despite this long waiting time, suggesting that the
structure is essentially frozen at these $T$ and $t_w$. This changes with
increasing $T$, curves for $T=0.25$ and $T=0.3$, where the amplitude of
$S(q)$ at low $q$ decreases rapidly, which shows that the large scale
structure starts to change significantly. The structure at intermediate
and large wave-vectors is, however, much less affected by this increase
of $T$. Thus we can conclude that the large scale structure of the network
has broken up, whereas the local structure is still relatively `intact' as
compared to  the initial structure of the relaxed cluster. As temperature
is increased even further the structure factor becomes essentially flat,
i.e.  the network structure of the system has broken up completely into
a gas of particles and small clusters.

\begin{figure}
\includegraphics[scale=1.0, angle=0]{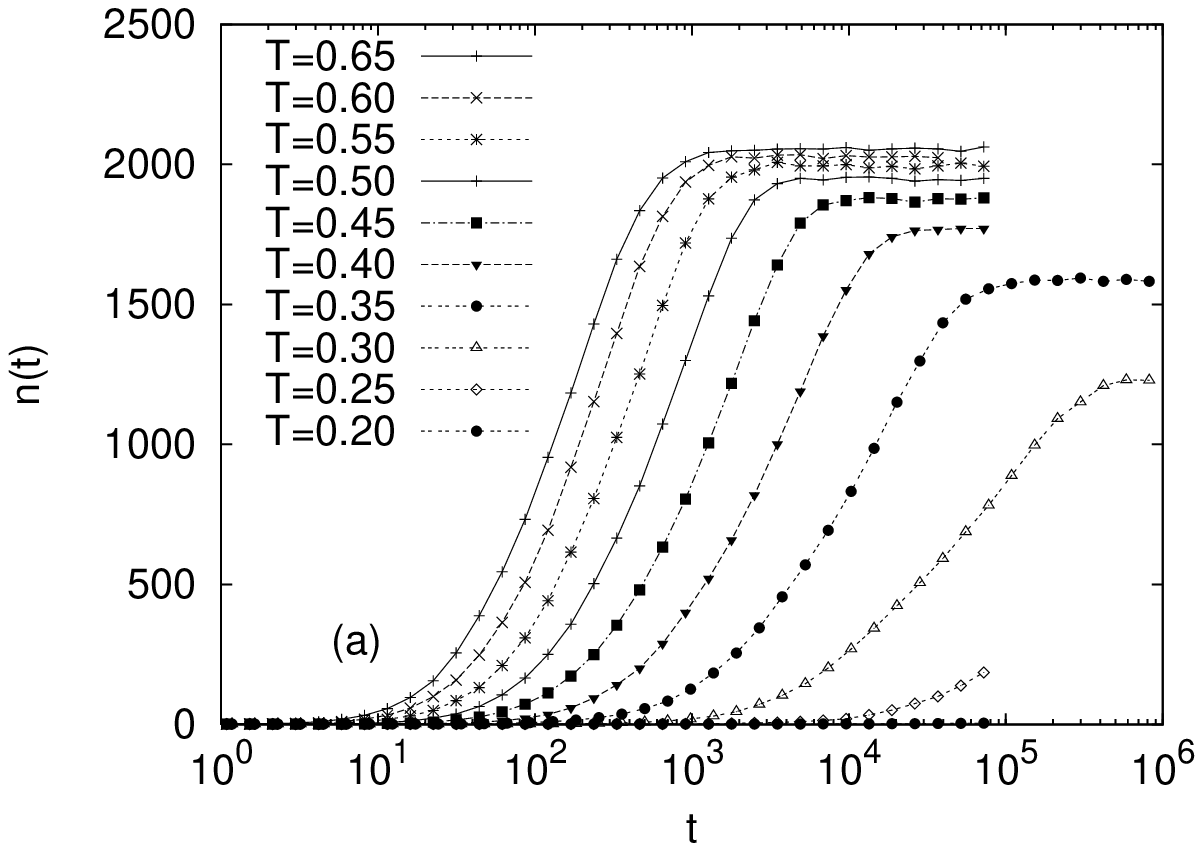}
\includegraphics[scale=1.0, angle=0]{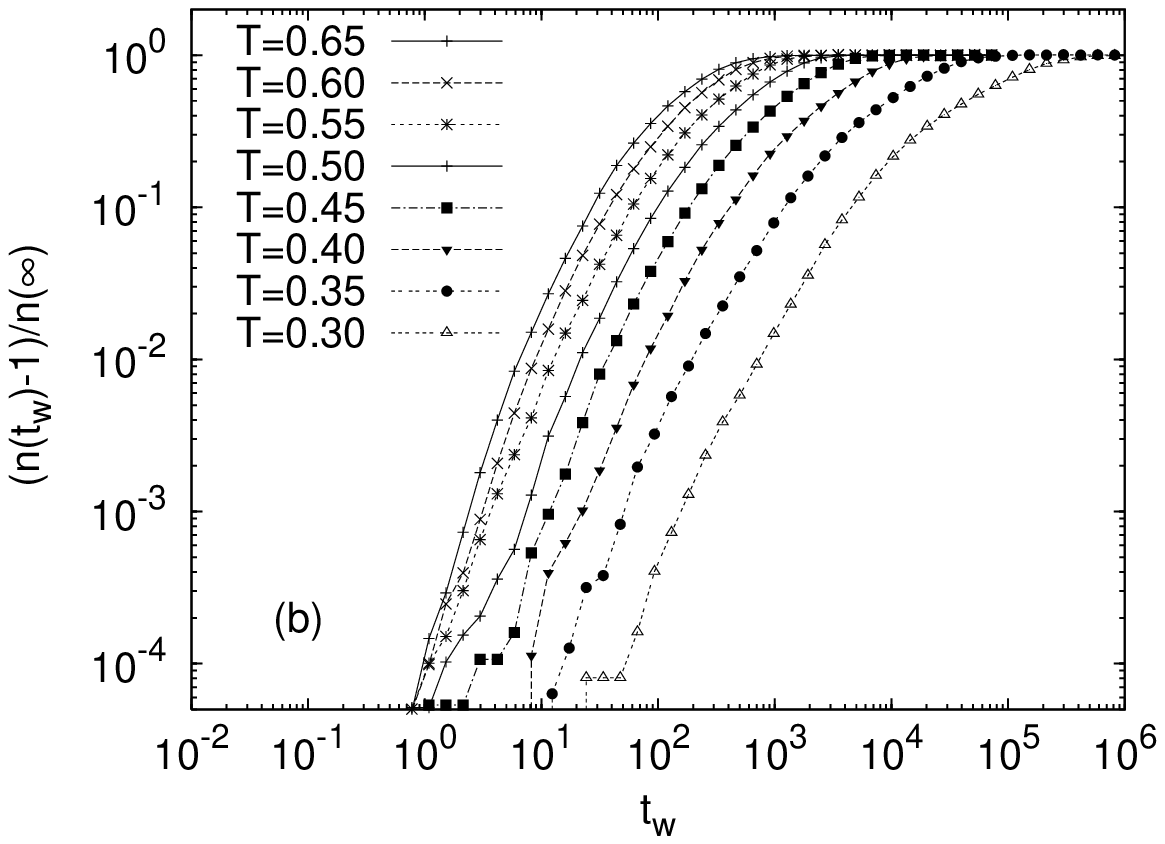}
\caption{
a) Number of clusters vs. waiting time, illustrating the breakup of
a single, percolating cluster, for various temperatures. b) The same
quantity as in a) but now normalized such that at $t_w\to \infty$ all
the curves converge to 1.0.
}
\label{Fig_Structure_NumberOfClusters}
\end{figure}

\paragraph{Cluster statistics}
The aging of the structure can also be analyzed by considering $n(t_w)$,
the number of clusters in the system as a function of the waiting time,
as shown in Fig.~\ref{Fig_Structure_NumberOfClusters} for medium
and high temperatures ($T\ge0.20$). For $t_w=0$ we have a single
(percolating) cluster. Then $n(t_w)$ increases with the waiting time
$t_w$, and the associated time-scale strongly depends on temperature,
see Fig.~\ref{Fig_Structure_NumberOfClusters}a.  Finally the number of
clusters  $n(t_w)$ saturates for large $t_w$, showing that the system
has completely relaxed, and the saturation value of $n(t_w)$ indicates
the number of clusters in equilibrium. Note that the saturation value
$n(t_w \to \infty)$ depends on temperature, since  most of the particles
are isolated at high $T$ whereas they prefer to form clusters at low $T$.

Figure~\ref{Fig_Structure_NumberOfClusters}a suggests that the shape
of $n(t_w)$ is independent of temperature, such that the only effect of
temperature would be a change in the relaxation time.  In order to test
this we consider the number of clusters (minus one), normalized by its
long time limit:

\begin{equation}
\frac{n(t_w)-1}{n(t_w \to \infty)}
\quad.
\end{equation}

\noindent
Note that this is of course only possible for those temperatures for which
there is a long-time plateau in Fig.\ref{Fig_Structure_NumberOfClusters}a,
i.e. for $T\geq 0.3$.

This quantity is shown as a function of $t_w$ in
Fig.~\ref{Fig_Structure_NumberOfClusters}b, confirming that the
shape of the normalized curves is indeed only a very weak function
of $T$. Furthermore we can conclude that the initial increase
of $(n(t_w)-1)/n(t_w \to \infty)$ is approximately a power-law
with an exponent close to 0.5. Although we do not have a complete
explanation for this particular exponent, it does demonstrate that
the initial increase is {\it not} exponential, as would be expected
for a process in which the initial cluster breaks up on a fixed
time-scale $\tau_{break}$ into sub-clusters, which in turn break up into
subclusters, etc., with a fixed time-scale for breakup.  On the contrary,
Fig.~\ref{Fig_Structure_NumberOfClusters}b suggests that clusters break
up in a heterogeneous manner regarding sizes and time scales.

\begin{figure}
\centering
\includegraphics[scale = 0.8]{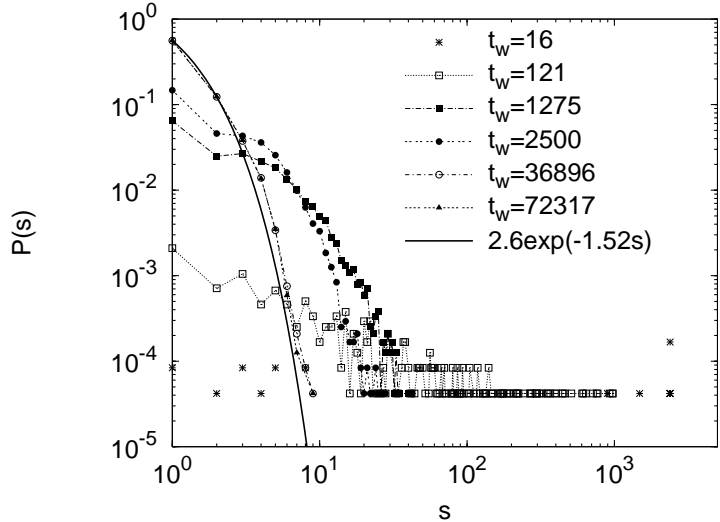}
\caption{
Cluster size distribution for high temperature ($T=0.40$), represented for
various ages. $P(s)$ is the probability of finding a cluster consisting of
$s$ particles in the system. Initially only one big cluster is present,
which decomposes progressively. The final regime is stationary (the last
two curves for large ages superpose), and are well described by 
an exponential distribution, solid line, as one would expect in a
gas of attractive particles. 
}
\label{Fig_PsT0.40}
\end{figure}

As mentioned above, the initial percolating cluster disintegrates into
smaller clusters and reaches an equilibrium for long times. Beyond the
asymptotic value of the number of clusters already mentioned above,
we can also analyze the breakup in terms of the evolution of the
cluster distribution $P(s)$ as a function of cluster size $s$. $P(s)$
depends on temperature (and at short and intermediate times also on
$t_w$). In Fig.~\ref{Fig_PsT0.40} we present $P(s)$ at   $T=0.4$ for
various waiting times $t_w$.  For $t_w=0$ the cluster size distribution
has a single peak at $s=N$, corresponding to the initial percolating
cluster. Even for small $t_w$ the distribution very quickly flattens
out, i.e. all cluster sizes become essentially equally likely (see curve
for $t_w=16$).  As $t_w$ is increased further the proportion of small
clusters increases quickly whereas that of large clusters diminishes.
For sufficiently long times, i.e. when the system has equilibrated, the
(asymptotic) distribution function is described well by an exponential
(solid line), since the particles just form random aggregation clusters
which follow Poisson statistics. Below we will make use of this result
when we discuss the relaxation dynamics of the system.

\section{Relaxation Dynamics}

After the discussion of the static structure of our model gel,
we now focus on its dynamical properties. As we will see below,
for intermediate and low temperatures, one can distinguish three time
regimes: (i) For $t_w$ much shorter than the time scale for the breakup
of the percolating cluster, the behavior is dominated by vibrational and
quasi-elastic effects. (ii) This is followed by a regime in which the
system ages strongly as its structure changes while the percolating
cluster breaks up. (iii) Finally, at very long times, the system has
reached equilibrium and we thus observe the relaxation dynamics of a
system consisting of small clusters. Although these three regimes are also
reflected in the time evolution of structural quantities, such as the
number of clusters shown in Fig.~\ref{Fig_Structure_NumberOfClusters}a,
we will show that the aging of the system affects the time correlation
functions much more strongly than the static quantities, in agreement
with previous investigations on glass-forming systems out of equilibrium
\cite{KobBarrat2000}. We will first analyze the average mean squared
displacement of the particles to extract information on the diffusive
(or non-diffusive) nature of the dynamics, before presenting a complete
study of the self-intermediate scattering function, and reporting on
the spatio-temporal characteristics of the relaxation processes.

\subsection{Mean squared displacement}

Since we are dealing with an out-of-equilibrium situation, the usual mean
squared displacement (MSD) has to be generalized by explicitly taking
into account the time at which the measurement is started. We thus define

\begin{equation}
\Delta^2 (t_w,t_w+\tau)  =  
\frac{1}{N} \sum_{i=1}^N \langle(\vec{r}_i(t_w+\tau) - \vec{r}_i(t_w))^2\rangle 
\qquad .
\end{equation} 

\begin{figure}
\centering
\includegraphics[scale=0.8]{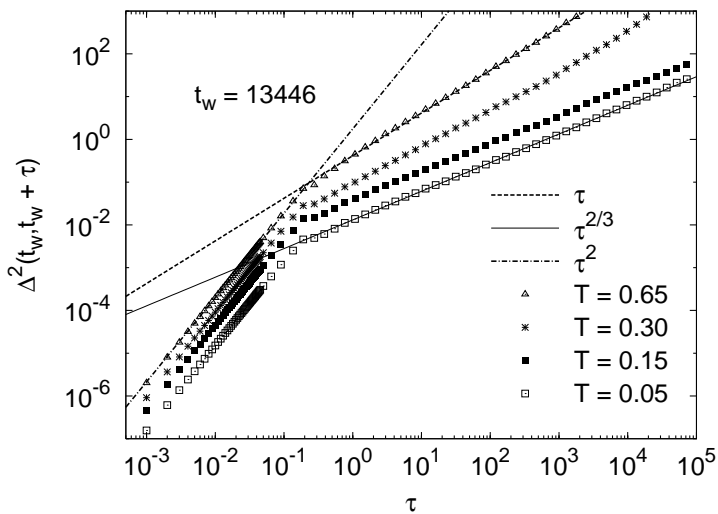}
\caption{
Mean squared displacement $\Delta^2 (t_w,t_w+\tau)$ of the particles as
a function of time $\tau$ for different temperatures, for $t_w=13446$. Also included are
power-laws found in the ballistic regime at small $\tau$ (dash-dotted
line), in the diffusive regime at large $\tau$ (dotted line), and in
the sub-diffusive regime (solid line).
}
\label{Fig_msd_tw13446.329}
\end{figure}

Since in aging glass-forming systems the relaxation dynamics has been
found to slow down with increasing $t_w$~\cite{leticia} we may expect
the same behavior in the present system. However, before addressing
this point it is useful to discuss the temperature dependence of
$\Delta^2(t_w,t_w+\tau)$ .  To this end we select a waiting time which is
relatively long, $t_w=13446$, and show in Fig.~\ref{Fig_msd_tw13446.329}
$\Delta^2(t_w,t_w+\tau)$ for different temperatures. At short times $\tau$
we find the ballistic behavior, $\Delta^2(t_w,t_w+\tau) \propto \tau^2$
(dashed-dotted line) as it is seen for all systems with Newtonian
dynamics. This ballistic regime ends at $\tau=0.25$, the time at which
the thermostat acts for the first time\footnote{In order to have a good
resolution for these values of $\tau$, a smaller time step, namely $h =
0.001$, has been chosen for this interval of the simulation.}.

\paragraph{Temperature dependence}
At high temperatures, $T=0.65$, this ballistic regime immediately
crosses over into a diffusive regime, $\Delta^2(t_w,t_w+\tau) \propto
\tau$ (dotted line in Fig.~\ref{Fig_msd_tw13446.329}). A look at
Fig.~\ref{Fig_Structure_NumberOfClusters} shows that for this temperature
and at this waiting time the initial percolating gel has already fallen
apart into many small clusters which have completely equilibrated, and
the observed $\Delta^2(t_w,t_w+\tau)$ is therefore just the MSD of the
disintegrated system consisting of clusters {\it in equilibrium}.

At low temperatures, on the other hand, the MSD shows a different
$\tau-$dependence once the ballistic regime is over. In this case the
MSD is sub-diffusive and can be well approximated by the power-law
$\Delta^2(t_w,t_w+\tau) \propto \tau^\alpha$ with an exponent $\alpha$
around 2/3 (solid line). Note that for sufficiently low temperatures $T$
this exponent appears to be independent of temperature: the curves for
$T=0.15$ and $T=0.05$ are parallel, and it is only for intermediate
temperatures, $T=0.3$, that the exponent of the sub-diffusive
regime exceeds 2/3.  We observe furthermore that at this $T$ this
sub-diffusive regime ends at around $\tau \approx 10^4$ where the
MSD starts to bend upwards, and from here on it grows proportionally
to $\tau$, which we rationalize as follows.  A comparison with Fig
\ref{Fig_Structure_NumberOfClusters} shows that, at this temperature,
the system starts to show significant aging effects on the time
scale of $\tau\approx 10^4$, i.e.  the number of clusters is already
relatively large although it has not yet reached the final (equilibrium)
distribution. It can be expected that many of these clusters are
relatively small and consequently move fast. Since the (average) MSD is
dominated by fast moving entities, it is therefore not surprising that
at $\tau \approx 10^4$ the MSD already shows diffusive behavior even
though the system is not in equilibrium yet.

Last but not least, we point out the absence of plateaus in the MSD
in Fig.~\ref{Fig_msd_tw13446.329} at intermediate time scales. Such
a plateau is a typical feature of most glass-forming systems at {\it
high} density~\cite{leticia}, where it is related to the so-called cage
effect, i.e. the temporary trapping of a particle by its surrounding
neighbors. In contrast to this the present gel has the structure of
a very open network, cf. the snapshot in Fig.~\ref{Fig_cluster2}.
Whereas in this system the cage effect may indeed be expected to be
present along the chains, we may conclude that the particle movements
from transverse motion, ie. orthogonal to the chains, is sufficiently
strong as to mask the plateau in the MSD, in agreement with previous
findings~\cite{delgado,zaccarellireview,LengthRelaxation_DelGado_Kob}.

\begin{figure}
\centering
\includegraphics[scale= 0.8]{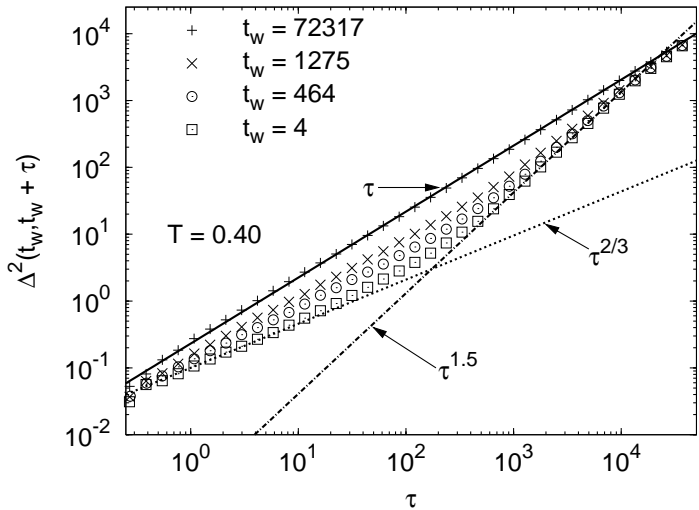}
\caption{
Mean squared displacement at $T = 0.40$ for different waiting times $t_w$.
Note that for reasons of clarity we do not show the ballistic regime
present at very small times ($\tau < 0.25$). Also included are power-law
fits with exponent 1.0, characterizing the diffusive regime seen for
$t_w=72317$, as well as with exponents 2/3 and 1.5 characterizing the
MSD at small $t_w$ at short and long times $\tau$, respectively.
}
\label{Fig_msdT0.40}
\end{figure}

\paragraph{Waiting time dependence} 
We now turn to the waiting-time dependence of the MSD. In
Fig.~\ref{Fig_msdT0.40} we show the $\tau-$dependence of
$\Delta^2(t_w,t_w+\tau)$ for different waiting times $t_w$,
all at a relatively high temperature of $T=0.4$, allowing us to
study the behavior of $\Delta^2$ in the aging regime as well as in
equilibrium (see Fig.~\ref{Fig_Structure_NumberOfClusters}). For short
waiting times, $t_w=4$, we see that after the ballistic regime (not
shown in the figure) $\Delta^2$ shows a $\tau-$dependence which is
compatible with $\tau^{2/3}$, i.e. the same power-law which we have
found for the MSD at {\it low} temperatures and long waiting times
(see Fig.~\ref{Fig_msd_tw13446.329}). Thus we can conclude that this
sub-diffusive behavior is not only limited to low temperatures, but can
be observed at all temperatures in the time window which corresponds to
the onset of aging. In contrast to this for intermediate times $\tau$
the MSD is super-diffusive, and compatible with a power-law with exponent
$\alpha\approx 1.5$. This behavior arises in the time regime in which
the structure of the system changes rapidly, leading to the complete
breakup of the gel (see Fig.~\ref{Fig_Structure_NumberOfClusters}). At
even longer times the MSD shows simple diffusive behavior, i.e. $\Delta^2
\propto \tau$.

For a larger waiting time, $t_w=464$, the MSD no longer shows the
intermediate $\tau^{2/3}$ behavior seen for intermediate times;
instead we observe a power-law with an exponent between 2/3 and 1.0.
On the other hand, the long time behavior is again compatible with a
$\tau^{3/2}$ behavior for the MSD. Thus we can conclude that for this
waiting time the initial relaxation dynamics is faster than that for short
waiting times, and once the initial structure of the system has changed
significantly it once more shows the same super-diffusive dynamics  as
for small $t_w$. Finally, for a long waiting time, $t_w=72317$, the
MSD enters the diffusive behavior right after the initial ballistic
regime, which is expected for a system in equilibrium at long times.
In summary, we thus observe that, after the initial ballistic regime,
the MSD of the system shows a power-law with exponent of roughly 2/3 at
an intermediate time window, as long as the gel has not yet undergone
major restructuring with respect to its initial state.  At intermediate
times, as the structure starts to change significantly, the MSD shows
a super-diffusive power-law with an exponent around 3/2.  Finally,
for very long times, the MSD shows the linear $\tau-$dependence found
in equilibrium systems.

\begin{figure}
\centering
\includegraphics[scale = 0.8]{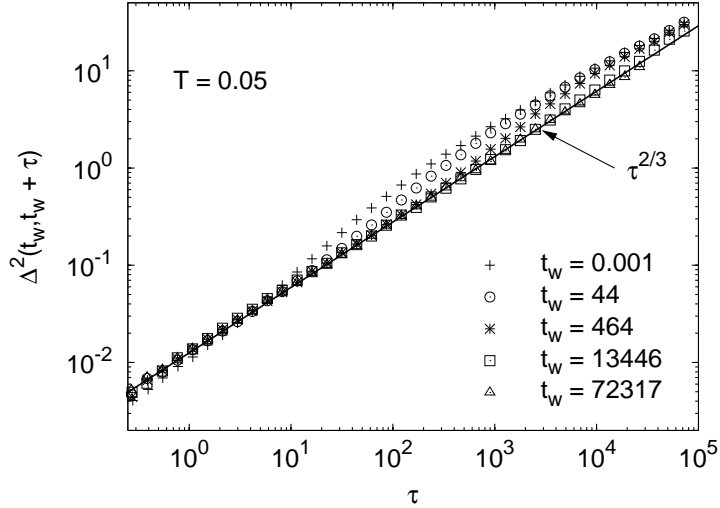}
\caption{
Mean squared displacement at $T= 0.05$ for different waiting times $t_w$.
Note that for reasons of clarity we do not show the ballistic regime
present at very small times ($\tau < 0.25$). Also included is a power-law
fit to the MSD for large $t_w$ with an exponent 2/3.
}
\label{Fig_msdT0.05}
\end{figure}

For the sake of completeness we also show the $\tau-$dependence of
the MSD at low temperature ($T=0.05$) and different waiting times,
see Fig.~\ref{Fig_msdT0.05}. For large waiting times we again find a
power-law with an exponent around 2/3. However, for smaller waiting
times, $t_w \leq 464$, the MSD shows a small (positive) deviation from
this power-law at intermediate times.  The reason for this is probably
the fact that at $t_w=0$ the structure is under weak tension (see the
discussion in the context of Fig.~\ref{Fig_Structure_Initial_RDF}), so
that the resulting forces lead to a somewhat faster relaxation dynamics
than what is observed at large waiting times, for which the tension has
already relaxed.

\paragraph{Dynamics of particle chains}
We devote the following paragraphs to the discussion of the mechanisms
which may give rise to a power-law dependence of the MSD with an
exponent around 2/3, which we have pointed out in various time windows
(Figs.~\ref{Fig_msdT0.40} and \ref{Fig_msdT0.05}). In particular,
we explore the role of the dynamics of particle chains forming
the gel network.  We recall that, at short and intermediate times,
the dynamics of the nodes in the gel network is much slower than the
transverse fluctuations of the chains. We may thus hope to understand
the anomalous $\tau-$dependence of the MSD by considering ``clamped''
individual chains, i.e. chains with endpoints that are fixed in space.

\begin{figure}
\centering
\includegraphics[scale= 0.8]{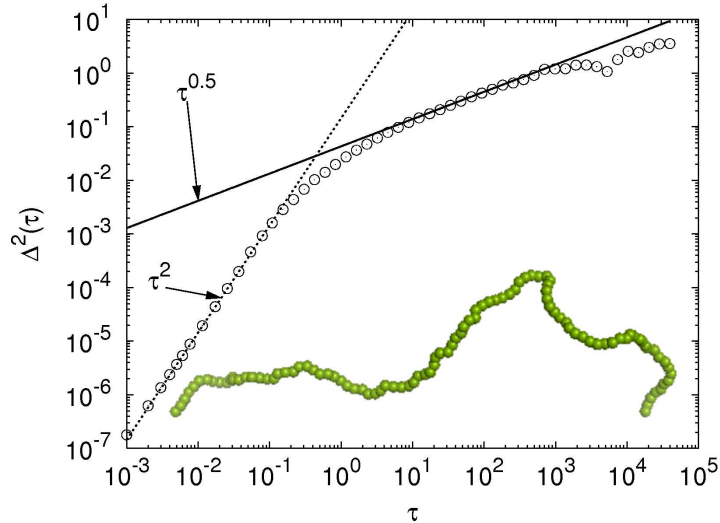}
\caption{
Mean squared displacement  $\Delta^2 (t_w,t_w+\tau)$ of a single chain
whose endpoints are clamped at a distance $L = R_{min}N$. $T=0.05$ and
$N=100$. The solid lines are fits with power-laws $\tau^2$ for short
times and $\tau^{0.5}$ at intermediate times. Note that the chains
have been equilibrated before starting the calculation of the MSD and
hence $\Delta^2$ does not depend on $t_w$.  Inset: Snapshot of a clamped
equilibrated chain of $N=100$ particles with length $L = 0.6 R_{min}N$,
obtained at $T=0.05$.
}
\label{Fig_msd_L1linear100}
\end{figure}

To this end we have performed MD simulations of isolated chains of
particles.  As starting configurations we used individual straight
linear chains of $N=100$ particles placed at their preferential distance
$R_{min}\approx 1.29$, which thus have a total length of $L=R_{min}N$.
Apart from holding the chain ends fixed, all other simulation details
(time step, heat bath, etc.) were identical to those used for simulating
the entire gel. The chains were thermalized at $T=0.05$ before starting
to acquire data for the MSD. In order to improve the statistics of the
results we have done 10 independent runs.

In Fig.~\ref{Fig_msd_L1linear100} we show $\Delta^2(\tau)$ as a function
of $\tau$ (note that, since this is a system in equilibrium, the MSD
does not depend on the waiting time, and we therefore omit the variable
$t_w$). At short times we have $\Delta^2(\tau) \propto \tau^2$ (ballistic
regime), whereas for long times  $\Delta^2(\tau) \propto \tau^{1/2}$,
as is expected for a Rouse chain~\cite{doi_and_edwards}. For very long
times the MSD saturates, since the chain is clamped at its ends which
in turn limits the displacement of particles in the chain to a certain
distance which  depends on $L$, $N$, and $T$. The small oscillations
seen at $\tau \approx 10^4$ are also related to the clamping through
fixed endpoints, since they are due to the excitation of the fundamental
mode of the chain.  Last but not least we point out that the mentioned
Rouse behavior is in fact not expected to be observable for chains
that are not very long, since it is known that for {\it free} short
chains the exponent 0.5 in the power-law is replaced by an exponent
around 0.6-0.7~\cite{baschnagelchong}. Since in our gel the length of
the connecting chains is not very large, see Fig.~\ref{Fig_cluster2},
it can be speculated that the exponent 2/3 found at short waiting times
is related to the relaxation dynamics of such (short) chains. However,
below we will discuss other mechanisms which potentially explain the
observed behavior.

\begin{figure}
\centering
\includegraphics[scale= 0.8]{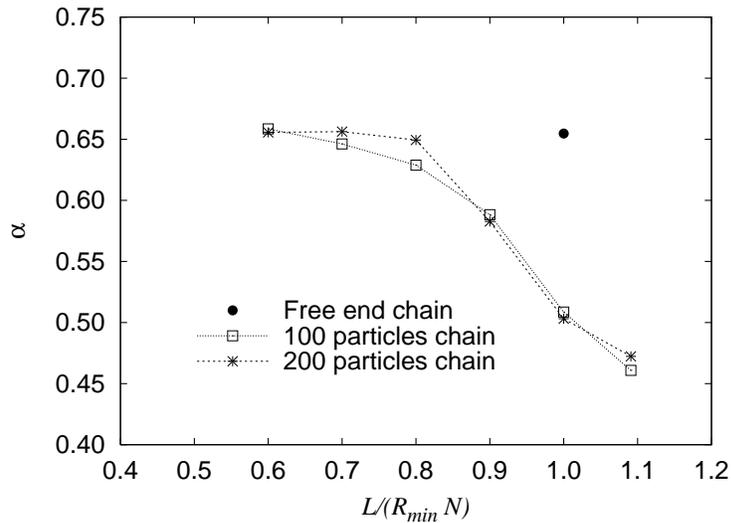}
\caption{
The exponent $\alpha$ defined via $\Delta^2 (t_w,t_w+\tau)\propto
\tau^{\alpha}$ as a function of the end-to-end distance of a clamped
chain.  $T=0.05$. See text for a detailed description of how the
initial configurations have been prepared. Note that the chain has
been equilibrated and hence the $\Delta^2$ does not depend on $t_w$.
The point labeled ``Free end chain'' corresponds to a chain in which
only one end was fixed, i.e. the value of the abscissa is irrelevant.
}
\label{Fig_alfa_vs_LN}
\end{figure}

The chains considered so far were not under tension, i.e. the distance
between the fixed ends was given by $L=R_{min}N$. In reality, however,
we have found the chains of the network to be under weak tension (see
the discussion in the context of Fig.~\ref{Fig_Structure_Initial_RDF}),
and it is therefore of interest to determine the  extent to which
the presence of such a tension modifies the relaxation dynamics. We
have therefore performed simulations in which the distance between
the end points of the chain were fixed at $L$ which differed from
$N \,R_{min}$.  A typical snapshot of such a chain is shown in the
inset of Fig.~\ref{Fig_msd_L1linear100}. For all cases we have found
that at intermediate and long times the MSD is given by a power-law
with an exponent $\alpha$ (and for very long times the MSD of course
saturates). In Fig.~\ref{Fig_alfa_vs_LN} we show the $L-$dependence of
this exponent and we see that for strongly compressed chains, $L< 0.7 <
R_{min}$, $\alpha$ is around 2/3. This limiting value can be rationalized
by the fact that most of the monomers of a significantly compressed
chain will not really feel that the ends of the chain are held fixed,
and therefore the MSD will behave very similarly to a semi-free chain. In
order to determine the value of $\alpha$ for such a semi-free chain,
we have run a simulation in which only one end of the chain was kept
fixed and, see Fig.~\ref{Fig_alfa_vs_LN}, we find that such a chain
indeed gives rise to an exponent $\alpha=2/3$.

If the distance between the ends of the chain is increased to $L=
R_{min}N$, on finds the exponent $\alpha=0.5$, in agreement with the
result shown in Fig.~\ref{Fig_msd_L1linear100}. If the chain is extended
even further, $L> R_{min}N$, the exponent becomes smaller than 0.5. Thus
we can conclude that a chain which is not clamped at its equilibrium
length shows a relaxation dynamic which differs from Rouse chain dynamics.

In order to check to what extent these results depend on the number of
monomers in the chain we have included in  Fig.~\ref{Fig_alfa_vs_LN} also
the exponents for a chain of length $N=200$. We see that this dependence
is not strong, in that the two curves are essentially identical for
chains at their equilibrium length, stretched chains or significantly
compressed chains. The only appreciable difference arises for slightly
compressed chains ($L/NR_{min}$ somewhat smaller than 1.0), for which
the exponent appears to increase with chain length.  This effect is due
to the fact that a longer chain will allow the particles far away from
the ends to undergo a relaxation dynamics which is less constrained than
in shorter chain, i.e. the dynamics is more similar to a free chain,
for which the exponent is 2/3.

\begin{figure}
\centering
\includegraphics[scale= 0.8]{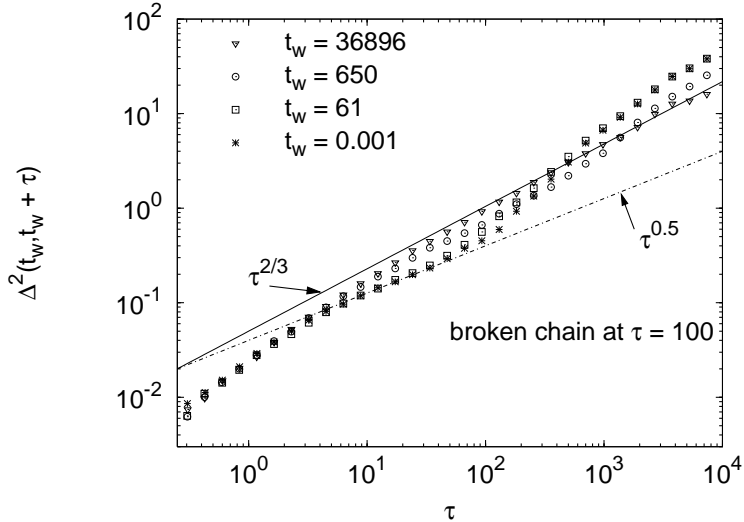}
\caption{
Mean squared displacement  $\Delta^2 (t_w,t_w+\tau)$ for a single
clamped chain of $N=100$ particles, obtained  at $T = 0.05$.  After having been
equilibrated, the chain, keeping its endpoints fixed, is artificially
cut in the middle at time $\tau=100$. Note that the ballistic part
is not shown.
}
\label{Fig_msd_break}
\end{figure}

In addition to the clamped chains we have also considered chains that
break up, as we must expect to happen naturally in the gel structure,
in order to study the contribution of such chain-breaking to aging.
For this we have used the equilibrated chains discussed above and started
an equilibrium simulation at time $t_0$. At time $t_0+100$ the chains were
then artificially broken in the middle and we have followed their out of
equilibrium dynamics from this point on.  Figure~\ref{Fig_msd_break} shows
the MSD for different waiting times $t_w$.  The data for $t_w=0.001$ shows
a crossover, at short times, from the ballistic regime to the dynamics
shown in Fig.~\ref{Fig_msd_L1linear100} which follow the $\tau^{0.5}$
power-law.  At $\tau=100$, the time at which the chain is broken, this
regime is interrupted and the MSD shows a super-diffusive dynamics
before it appears to slowly approach an asymptotic behavior. This
asymptotic long-time regime is more readily deduced from the data
for longer waiting times, see curve for $t_w=36896$, for which the
overall time which has elapsed since the chain was broken is much
larger. Consequently, although the behavior for small $\tau$ is as just
described, the asymptotic regime is attained over a much longer interval,
and we may deduce that the long-time chain relaxation of a broken chain
is compatible with a power-law $\Delta^2\propto \tau^{2/3}$. For even
longer times $\tau$, not easily accessible in our simulations, the MSD
must of course saturate again. We can therefore conclude that the out of
equilibrium relaxation of a chain following breakage obeys a power-law
with exponent 2/3. Consequently, chain breakage constitutes another
potential explanation fore the a $\tau-$dependence of MSD observed in
the model gel.

In summary one can conclude that the exponent 2/3 observed in the MSD
for the system at low temperatures and long waiting times can have at
least three origins: (i) rare ruptures of chains in the network, (ii)
large transverse fluctuations of floppy clamped chain-like filaments,
and (iii) fluctuations of dangling filaments. Currently we have not
attributed the anomalous diffusion effects to a dominant contribution
from either of these mechanism, which may in principle all coexist.

\subsection{Self-intermediate scattering function}

In order to obtain more detailed information on the relaxation
processes, in particular regarding the length-scale at which
they occur, we have investigated the self-intermediate scattering
function~\cite{hansenmacdonald86} generalized to a two-time quantity:

\begin{equation}
F_s(q,t_w,t_w+\tau) = \frac{1}{N} \sum^N_{j=1} 
\langle \exp[i\vec{q} \cdot (\vec{r}_j(t_w+\tau)-\vec{r}_j(t_w))] \rangle
\quad .
\label{eq5}
\end{equation}

\noindent
Here we have assumed that the system is isotropic and therefore $F_s$
depends only on $q=|\vec{q}|$. Recall that this function can be measured
experimentally  via scattering techniques~\cite{Pine2000}, and therefore
it is not only of theoretical interest but allows also to compare the
results from simulation with experimental data.

Since the relaxation dynamics at high temperature (where we have
interrupted aging~\cite{leticia}) is very different from the one at low
temperatures, we discuss them in two separate subsections.

\subsubsection{High temperature regime}

\begin{figure}
\centering
\includegraphics[scale=0.8]{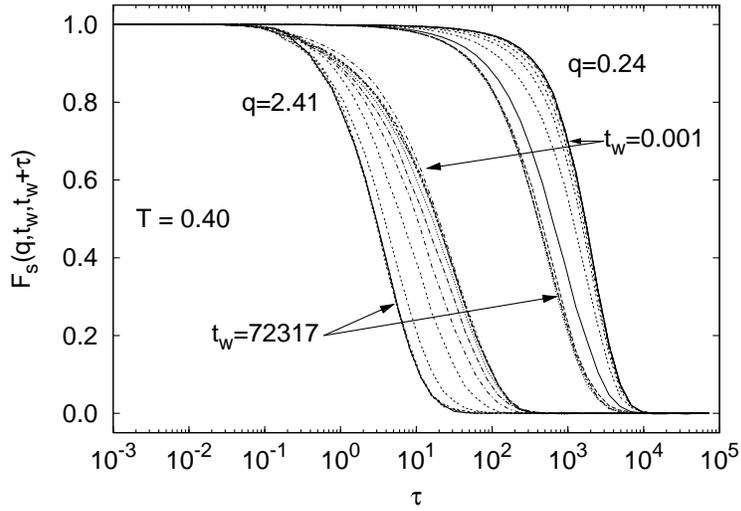}
\caption{
Time dependence of the generalized self-intermediate scattering function
$F_s(q,t_w, t_w+\tau)$ at a high temperature ($T=0.4$). The waiting
times are $t_w$ = 0.001, 4, 16, 44, 86, 237, 464, 1275, 4900, 13446,
26354, 51655, and 72317. Note that the relaxation time decreases with
increasing age.
}
\label{Fig_Fs_vs_tau_T0.40}
\end{figure}

In Fig.~\ref{Fig_Fs_vs_tau_T0.40} we show the $\tau-$dependence of
$F_s(q,t_w,t_w+\tau)$ for $T=0.40$, i.e. a temperature at which we can
see strong aging of the system, but which is still sufficiently high
to allow reaching equilibrium within the time span of the simulations,
see Fig.~\ref{Fig_Structure_NumberOfClusters} (higher temperatures give
a qualitatively similar relaxation dynamics).  Data are shown for two
wave-vectors: $q=0.24$, corresponding to a distance which is about 20
times larger than the typical nearest-neighbor distance (first peak
in the structure factor, see Fig.~\ref{Fig_Structure_SSF_Evolution}),
and $q=2.41$, corresponding to 2-3 particle diameters.  For each
of these wave-vectors we show several curves, which correspond to
different waiting times. One sees immediately that the shape of the
curves as well as the associated relaxation time depends on the age
of the system, and in the following we will discuss this dependence in
more detail. Even without any detailed analysis it is, however, evident
that  the relaxation dynamics accelerates with the sample age, as is
expected from the picture of an initially percolating structure which
disintegrates through clusters breaking up into ever smaller clusters,
which in turn break up themselves etc.  Note that this $t_w-$dependence
is different from the one found in {\it dense} aging glass-formers,
for which the relaxation time {\it increases} with age~\cite{leticia}.

In order to discuss the $q$ and $t_w$ dependence of the
relaxation dynamics we have fitted $F_s(q,t_w,t_w+\tau)$ with a
Kohlrausch-Williams-Watts (KWW) function, $A\exp(-(\tau/\tau_{f})^\beta)$,
using the time range in which the time correlation function has fallen
below 0.8. Although there is no theoretical justification for this
functional form, we have found that the fits are indeed very good, and
consequently the parameters $\tau_f(q,t_w)$ and $\beta(q,t_w)$ can be used
to characterize the relaxation time and the stretching of the dynamics.

\begin{figure}
\centering
\includegraphics[scale= 0.8]{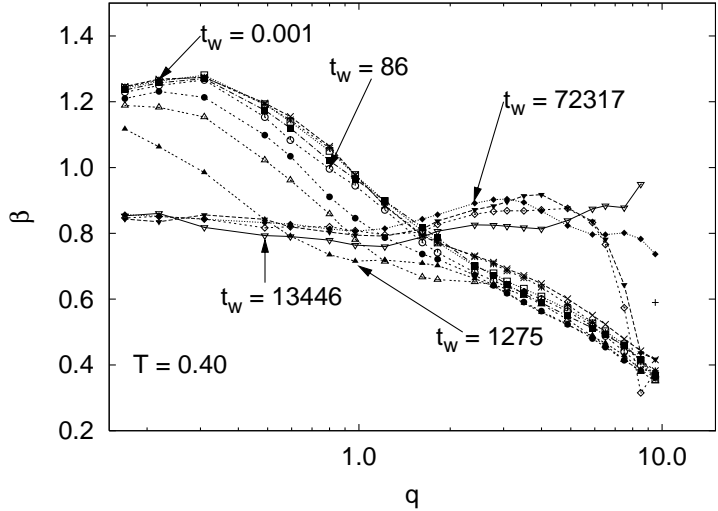}
\caption{
Wave-vector dependence of the KWW exponent $\beta$ at a high temperature
($T=0.4$). The waiting times are $t_w$ = 0.001, 4, 16, 44, 86, 237, 464,
1275, 4900, 13446, 26354, 51655, and 72317. Note that at small and large
$t_w$ the curves depend only weakly on $t_w$ whereas a significant
$t_w-$dependence is seen at intermediate waiting times.
}
\label{Fig_m_vs_qT0.40}
\end{figure}

\paragraph{Stretching exponent}
In Fig.~\ref{Fig_m_vs_qT0.40} we show the wave-vector dependence of the
KWW exponent $\beta$ for different waiting times. At short waiting times,
$t_w\leq 86$, the curves depend only weakly on $t_w$, since on this time
scale the gel ages only weakly. For intermediate times, however, $\beta$
depends strongly on $t_w$, since the system ages significantly in this
regime. Finally, for large $t_w$, the different curves fall onto a master
curve which describes the equilibrium dynamics of the system. Note
that at small waiting times the KWW exponent $\beta$ depends quite
strongly on $q$: We have $\beta \approx 1.2$ at large length scales,
which corresponds to a so-called {\it compressed} exponential, whereas
$\beta \approx 0.4$ at small length scales, corresponding to a strongly
{\it stretched} exponential.  We attribute this $q-$dependence to the
fact that the network structure is still intact at short $t_w$, such that
the high temperature imposes strong deformations on the filaments, and
therefore strong tensions.  Such stresses ultimately lead to an elastic
relaxation at a large scale (small $q$), which superposes and dominates
diffusive relaxation.  Such a compressed exponential, with exponent
$\beta = 3/2$, has indeed been predicted from an internal stress model
\cite{cipelletti-2005-17,PhysRevLett.84.2275}. A more refined microscopic
approach \cite{Bouchaud_Pitard_Eur.Phys.J.E} has lead, under certain
conditions, to predictions of $\beta = 5/4$, which is indeed close to
the exponents represented in Fig.~\ref{Fig_m_vs_qT0.40} at small $q$,
corresponding to length-scales on which one may hope that concepts of
elasticity apply.

We can furthermore rationalize why $\beta$ is small for large wave-vectors
and short $t_w$, by recalling that the short-time structure of the system
must locally retain some frustration, since it has been created by  what
is essentially a steepest descent procedure in the potential energy. The
resulting local geometry is therefore very heterogeneous and includes
regions with high stress and others with low stress, and which thus will
relax on quite different time-scales. We may therefore expect the {\it
average} relaxation dynamics to be a function which extends over many
time scales and which can be well fitted with a KWW function with a {\it
small} $\beta$. For long times, on the other hand, these local stresses
have successively homogenized, and therefore the typical relaxation
times no longer cover a large time window: the KWW exponent increases,
in agreement with what is seen in Fig.~\ref{Fig_m_vs_qT0.40}.

Interestingly, a strong $q-$dependence of the exponent
$\beta$ has also been observed in experiments of aging
gels~\cite{faraday} as well as in simulations of gels in
equilibrium~\cite{LengthRelaxation_DelGado_Kob}. These experiments show
that, at small $q$, the stretching exponent $\beta$ is indeed larger than
unity and decreases as a function of the wave-vector. This dependence is
thus qualitatively similar to the one found in the present study. However,
in the simulations of Ref.~\cite{LengthRelaxation_DelGado_Kob} the
trend is opposite, i.e. where $\beta$ {\it increases} with $q$, and
consequently we have $\beta>1$ only at large wave-vectors. The reason
for this difference probably resides in the fact that in the latter
work the microscopic dynamics was entirely Newtonian, i.e. there was
no coupling to a heat bath as in the present simulation. Consequently
the particles could follow a ballistic motion at small length-scales,
whereas such a time dependence is not possible here due to the presence
of the thermostat (which mimics the solvent of a real system, in a
rough way).  Furthermore, caution is in order when drawing parallels with
the dynamics of a gel in {\it equilibrium}, such as the one studied in
Ref.~\cite{LengthRelaxation_DelGado_Kob}, since it is  very different
from the out-of-equilibrium dynamics considered here.

\begin{figure}
\centering
\includegraphics[scale = 0.8]{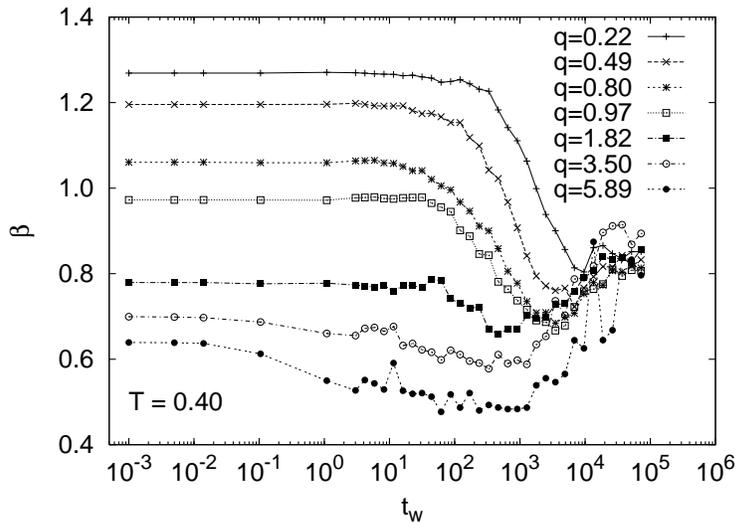}
\caption{
Aging time dependence of the Kohlrausch exponent $\beta$ at a high
temperature ($T=0.4$). The different curves correspond to different
wave-vectors.
}
\label{Fig_m_vs_twT0.40}
\end{figure}

\begin{figure}
\centering
\includegraphics[scale =0.8]{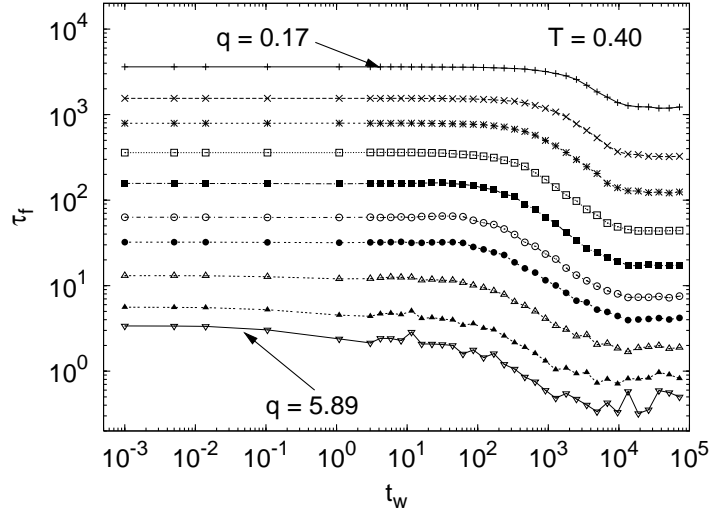}
\caption{
Relaxation time $\tau_f$ as a function of the age of the system for
different wave-vectors $q$ ($q=0.17, 0.31, 0.49, 0.80, 1.22, 1.82, 2.41,
3.50, 4.88,$ and $5.89$, top to bottom). The temperature is $T=0.4$.
}
\label{Fig_tf_vs_twT0.40}
\end{figure}

In Fig.~\ref{Fig_m_vs_twT0.40} we show the age dependence of the
Kohlrausch stretching exponent for different length-scales identified
through $q$, all at $T=0.4$, in order to understand how aging affects the
dynamics on different length-scales. For very short times the exponent
$\beta$ is constant for all $q$, i.e. the out-of-equilibrium situation of
the system does not yet affect its relaxation dynamics. For intermediate
times $\beta$ decreases, i.e. the dynamics becomes more stretched,
and from the data we see that the time window in which this decrease is
observed depends strongly on $q$. For large wave-vectors the increase of
the stretching is observed very early and followed by a zone of roughly
constant $\beta$ for intermediate times ($10^1 \leq t_w \leq 10^3$). For
small $q$, in contrast, the decrease of $\beta$ starts only at around
$t_w \approx 10^2$ and stops at $t_w \approx 10^4$. These results are
quite reasonable since the relaxation times strongly decrease with the
wave-vector $q$ (see Fig.~\ref{Fig_tf_vs_twT0.40}). Consequently, aging
will affect the relaxation dynamics  first on a small length-scale,
and only later on a large length-scale.

For even larger times the structure starts to break up and hence $\beta$
increases again. This acceleration is seen first at large wave-vectors,
since the system will equilibrate faster on small length-scales than
on large ones.  For very long waiting times the exponent $\beta$
becomes independent of $t_w$, since the system has equilibrated, and
converges to a value smaller than unity.  In the Appendix  we show
that this limiting value (around $\beta \approx 0.8$) can be directly
deduced from the equilibrium distribution of the cluster size shown
in Fig.~\ref{Fig_PsT0.40}.

\paragraph{Relaxation time}
We now turn to the relaxation time $\tau_f$ and discuss its dependence
on waiting time and wave-vector.  In Fig.~\ref{Fig_tf_vs_twT0.40} we
show $\tau_f$ as a function of $t_w$ for different wave-vectors. In
agreement with the results for the stretching exponent, we see three
different regimes: (i) For short waiting times $t_w$ the relaxation
time $\tau_f$ is constant. (ii) For intermediate $t_w$ the relaxation
time starts to decrease; note that the time at which this decrease is
seen depends significantly on $q$ and occurs much earlier for the short
length scales. (iii) Finally the relaxation time becomes independent
of $t_w$ since the system has reached equilibrium. The time at which
this happens is independent of $q$. Thus we can conclude that the time
window in which the aging occurs is large for large $q$ whereas it is
relatively narrow for small $q$, which is again in agreement with the
findings for the stretching exponent $\beta$.

\begin{figure}
\centering
\includegraphics[scale = 0.8]{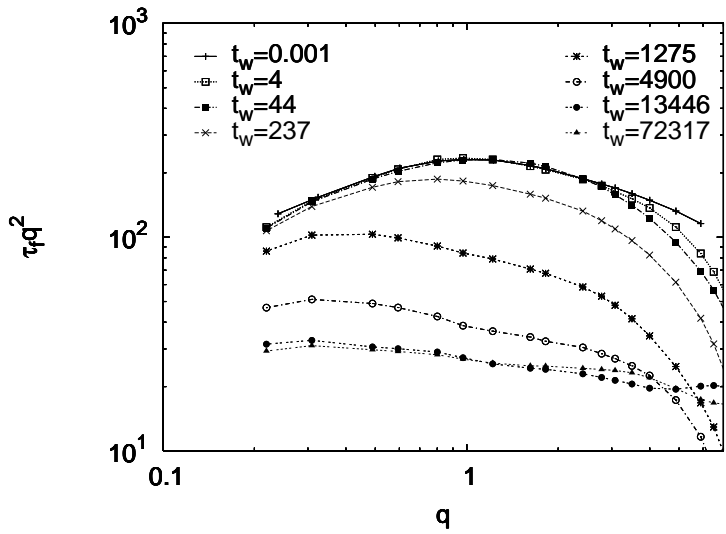}
\caption{
Wave-vector dependence of the relaxation time $\tau_f$ for different
waiting times $t_w$ at $T=0.4$. In order to check for the presence of
the hydrodynamic regime we plot $\tau_f q^2$.
}
\label{Fig_tf_vs_qT0.40}
\end{figure}

Finally we discuss the $q-$dependence of the relaxation time $\tau_f$. For
sufficiently long times we can expect hydrodynamic behavior, and we
therefore plot the rescaled relaxation time $\tau_fq^2$ as a function of
$q$, for different values of $t_w$, see Fig.~\ref{Fig_tf_vs_qT0.40}. The
data for short waiting times all collapse onto a master curve if
$q$ is small ($q \leq 3$), in agreement with the result shown in
Fig.~\ref{Fig_tf_vs_twT0.40}, whereas aging is observed only at the
largest values of $q$.  Note that we do not see a proper hydrodynamic
regime in this time window, since the rescaled curves are not flat even
at the smallest $q$. This is because the system is still in a gel phase,
and hence has a more complex dynamics.  For large $t_w$, however, we do
see the hydrodynamic behavior, i.e. $t_f \propto q^{-2}$ at small $q$, and
the hydrodynamic regime  extends to relatively large wave-vectors. This
is probably related to the fact that the asymptotic equilibrium system
consists of small clusters, which relax by a simple diffusion mechanism.

\subsubsection{Low temperature regime}

\begin{figure}
\centering
\includegraphics[scale = 0.8]{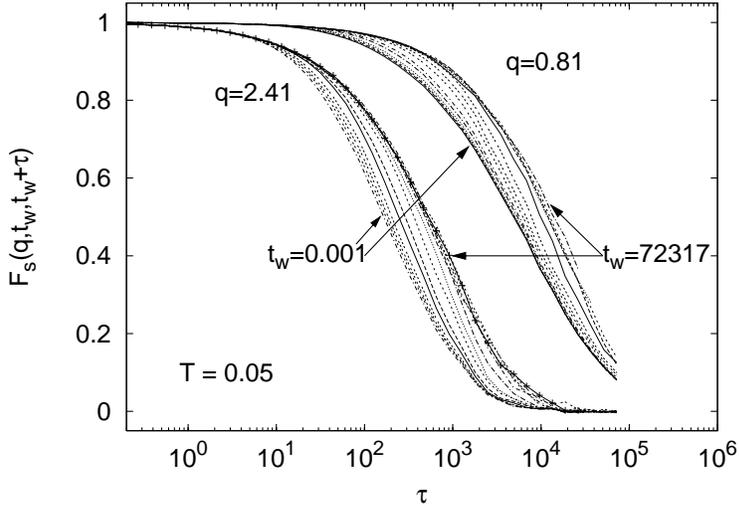}
\caption{
Time dependence of the intermediate scattering function
$F_s(q,t_w,t_w+\tau)$ at a low temperature ($T=0.05$). The wave-vectors
are $q=2.41$ and $q=0.81$. The different curves correspond to the waiting
times $t_w$ = 0.001, 4, 16, 44, 86, 237, 464, 1275, 4900, 13446, 26354,
51655, and 72317.
}
\label{Fig_Fs_vs_tau_T0.05}
\end{figure}

Having discussed the (interrupted) aging dynamics of the system at high
and intermediate temperatures, we now address the gel dynamics at low
temperatures, at which the system is not able to equilibrate within the
time span of the simulations.

In Fig.~\ref{Fig_Fs_vs_tau_T0.05} we show the time dependence of
$F_s(q,t_w,t_w+\tau)$ at $T=0.05$. Two different wave-vectors are
shown, and the different curves correspond to different waiting times
$t_w$. From this graph we see immediately that  the dynamics {\it
slows down} with increasing $t_w$, in contrast to what we have found
for intermediate and high $T$ (see Figs.~\ref{Fig_Fs_vs_tau_T0.40}
and \ref{Fig_tf_vs_twT0.40}).  Thus we can conclude that the aging
dynamics at low temperatures is very different from the one seen at
intermediate and high temperatures.  In particular, it is not possible
to interpret the low-temperature dynamics as a special case of the high
temperature dynamics at intermediate times, i.e. before the aging is
interrupted.  This difference can be understood by recalling that the
intermediate and high $T$ aging dynamics is dominated by the breakup of
the percolating cluster, whereas at low $T$ this cluster stays intact
within the  simulation time and aging is related to the rearranging of
its structure, see Fig.~\ref{Fig_Structure_NumberOfClusters}.

In order to discuss the details of the $q$ and $t_w-$dependence
of $F_s(q,t_w,t_w+\tau)$ we deduce a relaxation time $\tau_f$
and a stretching exponent $\beta$ by fitting the correlator to a
KWW-function. The resulting fit characterizes the shape of $F_s$
sufficiently well to justify using these parameters, provided we exclude
the smallest wave-vectors ($q\leq 0.5$); for these, the correlators
$F_s(q,t_w,t_w+\tau)$ do not decay significantly for the low temperatures
under consideration, and a KWW fit would be clearly inappropriate since
the fit parameters would be determined essentially from the short time
dynamics and not from the $\alpha-$relaxation.

\begin{figure}
\centering
\includegraphics[scale = 0.8]{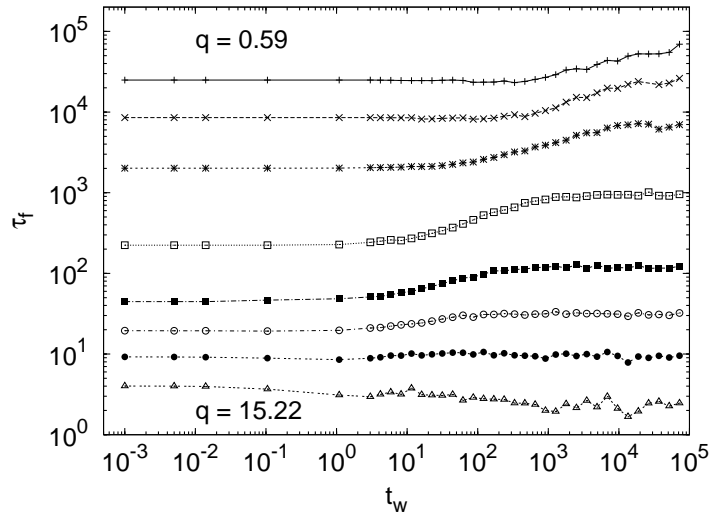}
\caption{
Waiting time dependence of the relaxation time at $T=0.05$. The different
curves correspond to the wave-vectors $q=0.59, 0.80, 1.22,  2.41, 
4.88, 7.50, 10.66$, and $15.22$ (from top to bottom).
}
\label{Fig_tf_vs_tw_T0_05}
\end{figure}

\paragraph{Relaxation time}
The waiting time dependence of the relaxation time $\tau_f$ is shown in
Fig.~\ref{Fig_tf_vs_tw_T0_05}, for various wave-vectors.  We see that
for all $q\leq 11$ the dynamics slows down with increasing age, whereas
a slight acceleration is observed for the highest $q$. The increase of
$\tau_f$ with age is the typical behavior found in many glassy systems
which are aging.  However, in those systems $\tau_f(t_w)$ is usually seen
to obey power-law dependence on $t_w$, or at least so whenever $t_w$
is sufficiently large~\cite{leticia}.  This is not really the case for
the system studied here, since from Fig.~\ref{Fig_tf_vs_tw_T0_05} we can
conclude that $\tau_f$ shows three regimes: (i) At short waiting times the
relaxation time is independent of $t_w$; the system has not yet had time
to react to its changing structure, in agreement with the $t_w-$dependence
of $\tau_f$ found at high $T$ (see Fig.~\ref{Fig_tf_vs_twT0.40}).
(ii) Subsequently aging starts to modify the structure; the stress
present in the initial cluster spreads, but does so without breaking
off smaller clusters (see Fig.~\ref{Fig_Structure_NumberOfClusters}b);
this second regime is therefore essentially an elastic relaxation which
allows the cluster to accommodate the interaction potential.  Note that
this regime lasts only for about two decades in time, before it crosses
over to (iii) a regime in which, for intermediate wave-vectors,  $\tau_f$
becomes independent of $t_w$.  From Fig.~\ref{Fig_tf_vs_tw_T0_05} we
see that the times at which the second regime starts and ends  depend
quite strongly on $q$, and that for the smallest wave-vector we are not
able to see the third regime at all within the time span accessible to
the simulation.  This is evidence that the stiffening of the structure,
which results in an increasing relaxation time, occurs on the local
scale before it affects the system on large length-scales.  Since the
simulated system is rather limited in size, this aging regime does not
last very long, and the relaxation times do not increase very much (by
less than a factor of five). Although we have not explicitly checked for
the presence of finite size effects, we are thus led to think that in
the thermodynamic limit the length of the second regime, as well as the
increase of $\tau_f$, must be much larger than the effects seen in our
small sample. This would be in agreement with the experimental findings
in real gels~\cite{cipelletti-2005-17}, for which the sample sizes are
much larger than the ones accessible to any simulation.

\begin{figure}
\centering
\includegraphics[scale = 0.8]{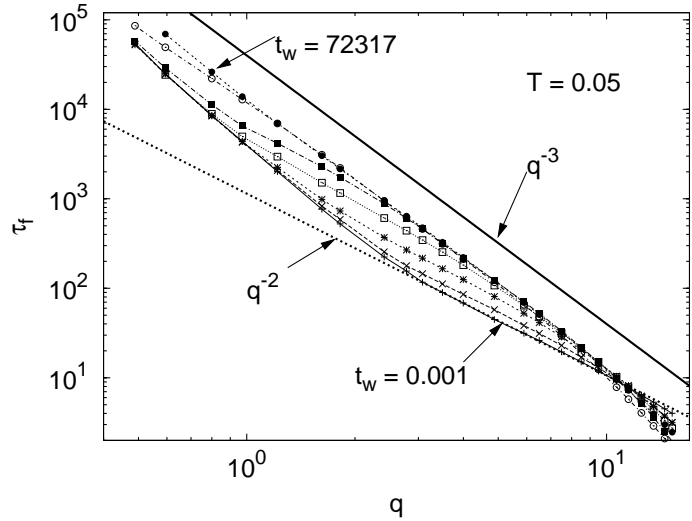}
\caption{
Relaxation time $\tau_f$ as a function of the wave-vector $q$ for
$T=0.05$. The different curves correspond to the waiting times $t_w =
0.001$, 8, 44, 237, 1275, 13446, and 72317.
}
\label{Fig_tf_vs_qT0.05}
\end{figure}

\begin{figure}
\centering
\includegraphics[scale = 0.8]{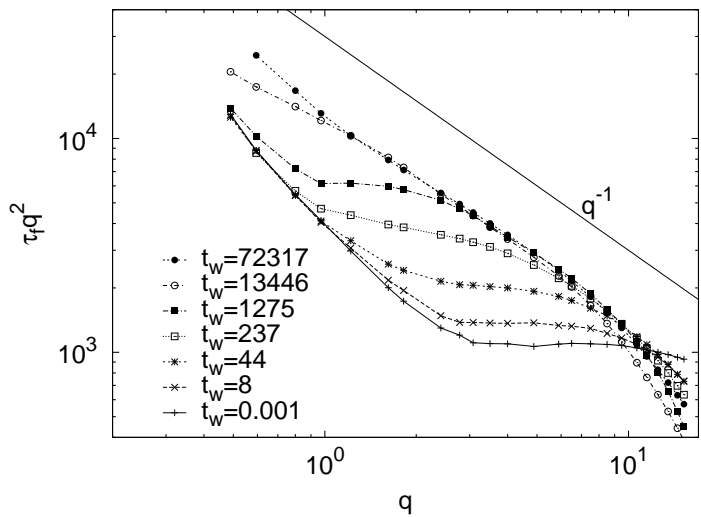}
\caption{
Relaxation time $\tau_f$ times $q^2$ as a function of the wave-vector $q$
for $T=0.05$. The different curves correspond to the waiting times $t_w
= 0.001$, 8, 44, 237, 1275, 13446, and 72317 (from bottom to top). The
solid curve correspond to a $q^{-3}-$dependence of $\tau_f$.
}
\label{Fig_tfq2_vs_qT0.05}
\end{figure}

Next we analyze the wave-vector dependence of the relaxation time
$\tau_f$, which is shown in Fig.~\ref{Fig_tf_vs_qT0.05} for several
ages $t_w$ ($0.001\leqslant t_w\leqslant 72317$).  We first consider
short waiting times.  For intermediate and large wave-vectors we find
a $q^{-2}$ scaling for $\tau_f(q)$, corresponding to a diffusive motion
(dotted line).  For small wave-vectors, $q\leq 2.0$, this crosses over
to a stronger dependence, and we find approximately $\tau_f(q) \propto
q^{-3}$ (solid line).  For long waiting times, on the other hand, we
find that $\tau_f(q)$ is given by a power-law $q^{-3}$ in essentially the
entire accessible $q$-range (small deviations arise only at the largest
$q$). Note that this $q-$dependence implies that the relaxation time on
a scale $r$ is proportional to $r^3$, and that for a given time scale
$\tau$ the particles will therefore reach a typical distance $r(\tau)
\propto \tau^{1/3}$. This implies that the mean squared displacement
increases like $\tau^{2/3}$, a result that is indeed in agreement with
the $\tau$-dependence we have found for the MSD for long waiting times
and low $T$, see Fig.~\ref{Fig_msd_tw13446.329}.

In order to analyze in more detail the diffusive behavior for
large wave-vectors $q$ and short waiting times, as identified
from Fig.~\ref{Fig_tf_vs_qT0.05}, we re-scale the data by plotting
$\tau_f  q^2$, see Fig. \ref{Fig_tfq2_vs_qT0.05}. This confirms that
the relaxation dynamics is indeed diffusive on intermediate and large
length-scales for short waiting times (bottom curves). For larger waiting
times $t_w$ the $q$-window for diffusion shifts to smaller wave-vectors
and shrinks, finally giving way to a $\tau_f \propto q^{-3}$ scaling
over the whole accessible $q-$range for the largest waiting times.
This graph thus nicely highlights that the relaxation dynamics strongly
depends on the wave-vector and on the age considered.

\begin{figure}
\centering
\includegraphics[scale = 0.8]{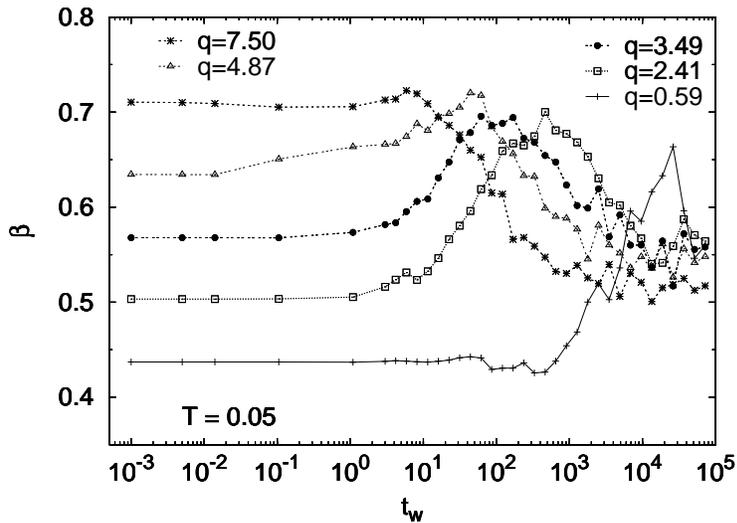}
\caption{
The  Kohlrausch exponent $\beta$ as a function of the age $t_w$.
$T=0.05$. The different curves correspond to different wave-vectors
$q$. Note that with increasing $q$ the maximum of $\beta$ moves toward
smaller values of $t_w$.
}
\label{Fig_m_vs_tw_T0.05}
\end{figure}

\begin{figure}
\centering
\includegraphics[scale = 0.8]{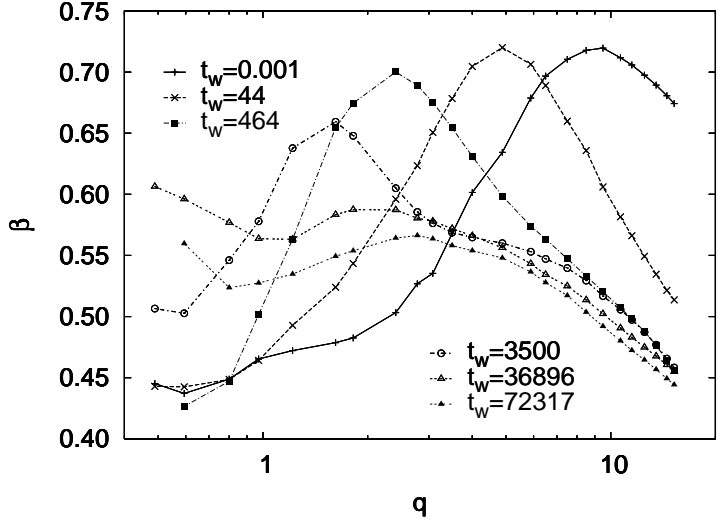}
\caption{
The  Kohlrausch exponent $\beta$ as a function of the wave-vector
$q$. $T=0.05$.  The different curves correspond to different ages $t_w$.
}
\label{Fig_m_vs_qT0.05}
\end{figure}

\paragraph{Kohlrausch exponent}
Finally we analyze how the Kohlrausch exponent $\beta$ depends on the
wave-vector and on the waiting time. In Fig.~\ref{Fig_m_vs_tw_T0.05} we
show $\beta$ as a function of $t_w$ for different values of $q$. First
of all we see that $\beta<1$ for all ages $t_w$ and all wave-vectors
$q$, i.e. the relaxation dynamics is always stretched. This finding
is in contrast to experimental and theoretical results for aging
gels, for which compressed exponentials ($\beta>1$)  have been reported
\cite{PhysRevLett.84.2275,cipelletti-2005-17,Bouchaud_Pitard_Eur.Phys.J.E,LengthRelaxation_DelGado_Kob}.
One possible explanation for this difference is the fact that the length
scales accessible here are much smaller than the ones considered in
the experiments or through theoretical arguments. In the cited studies
the compressed exponentials have been related to the presence of a local
rupture of the network and its subsequent elastic relaxation, which leads
to a ballistic motion.  An elastic relaxation here means that the part of
the gel which has undergone rupture is pulled by a constant force. Due
to the presence of the solvent the resulting motion is balanced by a
frictional force, and therefore the resulting motion is linear in time,
i.e. ballistic.  Since in our simulation the network does not break at
$T=0.05$ and on the time scales considered, this type of relaxation is not
present and hence $\beta<1$.  Let us point out that our results are {\it
not} in contradiction with the results from a computer simulation of a gel
{\it in equilibrium} mentioned above~\cite{LengthRelaxation_DelGado_Kob},
for which $\beta>1$ was found for large wave-vectors. Since the
simulations in Ref.~\cite{LengthRelaxation_DelGado_Kob} were conducted
with Newtonian dynamics (unlike our Brownian-like dynamics due to the
thermostat), they trivially give rise to a ballistic motion at small
length-scales, which in turn imply a compressed exponential relaxation
\cite{Bouchaud_Pitard_Eur.Phys.J.E}.

>From Fig.~\ref{Fig_m_vs_tw_T0.05} we can further deduce that the
stretching at small $t_w$ increases continuously with decreasing $q$
(as long as $q$ is not too large), which is evidence that the system
becomes increasingly heterogeneous on increasing length-scales. For all
values of $q$ we see that, as a function of $t_w$, the exponent $\beta$
is essentially constant at small $t_w$, then goes through a local maximum
before converging to a constant value for large $t_w$. Remarkably, this
constant of around 0.55 is independent of $q$. The age at which $\beta$
shows its local maximum corresponds roughly to the aging time at which
the relaxation time $\tau_f(t_w)$ shows an inflection point, i.e. a
strong $t_w-$dependence, see Fig.~\ref{Fig_tf_vs_tw_T0_05}. Thus we are
lead to conclude that in the time window in which the relaxation time
changes significantly (and which depends on the length-scale considered,
see Fig.~\ref{Fig_tf_vs_tw_T0_05}) the stretching is weaker than in
the regime where the system is not aging quickly.  Put differently,
rapid aging dynamics is associated with weak stretching, and is probably
linked to elastic rearrangements.

Finally we show the wave-vector dependence of $\beta$ for different
waiting times, see Fig.~\ref{Fig_m_vs_qT0.05}. For small $t_w$, the
dynamics is not very stretched at a length-scale corresponding to $q
\approx 10$ (pronounced maximum in $\beta(q)$), whereas the stretching
is much stronger for $q \approx 1$.  With increasing age of the system
the  maximum in $\beta(q)$ shifts to smaller wave-vectors, i.e. the
length-scale on which the system relaxes increases with age. For very
long waiting times, $\beta(q)$ appears to converge towards a master
function showing a pronounced decrease of $\beta$ at intermediate and
large $q$. This master curve is furthermore qualitatively similar
to the behavior seen for $\beta(q)$ at higher temperatures, see
Fig.~\ref{Fig_m_vs_qT0.40}, where we observe a local maximum around $q=2$,
and an increase for smaller $q$'s.  The complex shape of $\beta(q)$
seen at long times therefore reflects the relaxation dynamics of the
disordered elastic network, in which several relevant length-scales
interfere simultaneously.

\section{Conclusions}

In this article we have studied a system of particles designed to mimic
a physical gel. We started from an ad hoc fractal structure, constructed
from a modified DLCA algorithm and therefore known to reproduce the
structure of dilute, fractal colloidal gels. Although this approach is
clearly not inspired by thermodynamics -DCLA-DEF is a purely kinetic
algorithm- it has been shown to produce structures which successfully
reproduce several measurable quantities of gels, such as the density
dependence and the rigidity modulus \cite{JourNon-CrystSol_Ma_Jullien}.
It was by no means granted, however, that this way of making a gel-like
structure could also be a satisfactory starting point  regarding the
thermodynamics, left alone the (equilibrium or out-of-equilibrium)
dynamics, as we have shown to be the case.

In this study we have investigated the stability of the initial
DLCA structure as a physically reasonable particle pair potential is
imposed onto  the kinetic DLCA structure. Letting the system adapt
to this potential, via an initial zero-temperature relaxation, led to
significant changes in local ordering while leaving the fractal exponent
essentially unaffected.  The subsequent relaxation at a given temperature
$T$ evolves according to one of the following scenarios.

At high temperature ($T>0.2$), the gel is unstable and ultimately breaks
up into small clusters of particles.  At low temperature ($T<0.2$) the gel
is stable and ages slowly, i.e its aging slows down with observation time.

For high $T$, two regimes for the waiting time must be distinguished.
For small $t_w$, the self-intermediate scattering function $F_s$ is
a stretched exponential in time on small length-scales -reflecting
the frustrated local structure- and a compressed exponential on large
length-scales, where the high temperature imposes strong deformations,
such that elastic relaxation dominates. At large $t_w$, when the
initial structure has decomposed into small pieces, $F_s$ is a stretched
exponential with essentially a constant exponent $\beta=0.8$ for all
wave-vectors, which can be rationalized from the measured long-time
distribution of cluster sizes (see Appendix).

The low-temperature relaxation is rather complex. At large $t_w$, the mean
squared displacement behaves like $\tau^{2/3}$, a behavior which reflects
the dynamics of chain-like filaments connecting the nodes of the gel
network. We showed that the reason for this dynamics could be related to
a combination of large transverse fluctuations of floppy, clamped chains,
to fluctuations of dangling filaments and also to rare rupture events of
clamped chains. This behavior of the mean squared displacement translates,
in the study of relaxation times in Fourier space, as $\tau_f(q) \propto
q^{-3}$ for large $t_w$ and small wave-vector.  At low $T$, the exponent
$\beta$ of the self-scattering function is always smaller than unity and,
for small $t_w$, it decreases as $q$ decreases: Dynamical heterogeneity
increases with the length-scale.  For larger $t_w$, $\beta$ goes through
a maximum (corresponding to the timescale where the aging of the gel is
most pronounced). On this timescale, elastic rearrangements are probably
important, but do not dominate over heterogeneity.

In conclusion, we have investigated the properties of the system as a
function of temperature, both in terms of structure and dynamics, using
a particle  volume fraction comparable to experimental colloidal gels. We
have not studied the effect of the volume fractions, as could be done in
future work. From the present study, we have established a body of results
which are suitable for confrontation with experimental measurements. In
particular, to our knowledge, the breaking dynamics of a gel at high $T$
has not been reported experimentally. Concerning gels at low $T$, where
the gel structure is (meta)stable but ages, our study presents detailed
results concerning relaxation on a local length-scale. Experiments on soft
gels mostly  reveal compressed exponential relaxation (light scattering
experiments \cite{faraday,cipelletti-2005-17,PhysRevLett.84.2275}, and
X-ray scattering experiments \cite{ranjini,carbonblack}), and highlight
the importance of internal elastic stresses. In our model system this
behavior is seen only at high temperatures, implying that the relaxation
of internal stresses is still hard to tackle within numerical simulations
with Brownian dynamics.

Finally, we hope that our results will hint at a novel and more
refined interpretation of relaxation measurements in gels, in particular
concerning the fluctuations of particle filaments. To this end, deducing
the fluctuation parameter $\chi(q,T,t_w,\tau)$, which measures the
variance of the self-intermediate scattering function, from numerical
simulations would greatly help in understanding the underlying microscopic
and/or mesoscopic mechanisms.

\section{Appendix}
\label{appendix}

In this Appendix we show that the relaxation dynamics of the system at
intermediate and high temperatures and at long times can be calculated
directly from the measured cluster size distribution presented in
Fig.~\ref{Fig_PsT0.40}. For this we consider a cluster of particles
having size $s$. For times shorter than $\Delta_{therm}$, the period
for the thermostat, this cluster will essentially move on a straight
trajectory since $\Delta_{therm}$ is shorter than the mean free collision
time. The length of this ballistic trajectory is of the order of $\ell =
v \Delta_{therm}$, where $v$ is the velocity of the cluster. From the
equipartition theorem we can estimate $v$ to be given by

\begin{equation}
\frac{1}{2} sm v^2 = \frac{3}{2}k_BT \quad .
\label{eqa1}
\end{equation}

\noindent
Since after time $\Delta_{therm}$ all particles are coupled to the heat
bath, the velocities of all particles, and hence the one of the cluster,
are randomized. Hence the cluster will perform a random walk with mean step
size $\ell$. After a time $t$ the cluster will therefore have made a mean
squared displacement $R^2(t)$ given by

\begin{equation}
R^2(t)= \frac{t}{\Delta_{therm}} \ell^2 
= \frac{t}{\Delta_{therm}} v^2 \Delta_{therm}^2
= \frac{3  k_B T \Delta_{therm} t}{ms} \quad .
\label{eqa2}
\end{equation}

For a random walk the self-intermediate scattering function $F_s(q,t)$ is
given by~\cite{hansenmacdonald86}

\begin{equation}
F_s(q,t) = \exp(-q^2 R^2(t)/6) = \exp(-q^2 k_B T \Delta_{therm}t/2ms) \quad .
\label{eqa3}
\end{equation}

\begin{figure}
\centering
\includegraphics[scale = 0.8]{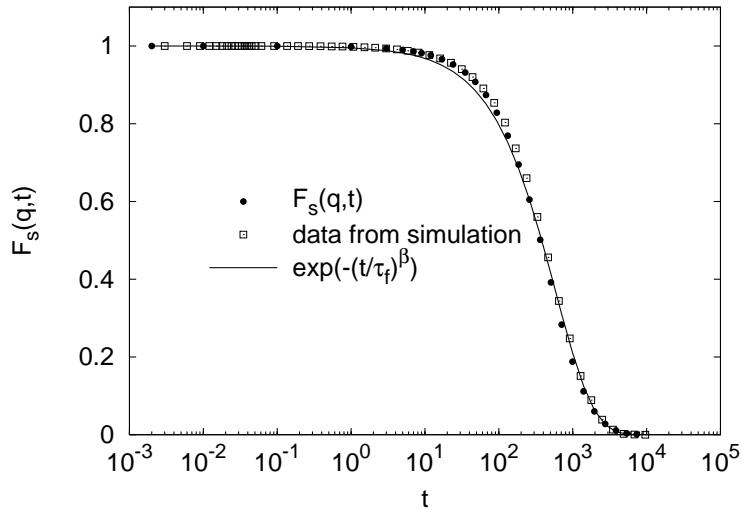}
\caption{
Time dependence of the self-intermediate scattering function at $T=0.4$ for
waiting times that are sufficiently large to allow the system to reach
equilibrium (squares). Also included is a theoretical curve for $F_s(q,t)$ as
calculated from the cluster size distribution using Eq.~(\ref{eqa4}) (filled
circles). Both functions can be fitted very well by a stretched exponential
with a parameter $\beta=0.84$ (solid line).
}
\label{Fig_beta_rmsT0.4}
\end{figure}

Thus if we have a system that is composed of a collection of clusters with a 
size distribution $P(s)$, the self-intermediate scattering function is
given by

\begin{equation}
F_s(q,t) = \frac{1}{\sum_{s=1}^\infty s P(s)} \sum_{s=1}^{\infty} s P(s) 
\exp(-q^2 k_B T \Delta_{therm} t/2ms) \quad .
\label{eqa4}
\end{equation}

Since we have seen that {\it in equilibrium} our system has an exponential
distribution of cluster sizes, see Fig.~\ref{Fig_PsT0.40}, we have

\begin{equation}
P(s)= \exp(-\alpha s) \qquad ,
\label{eqa5}
\end{equation}

\noindent
where the parameter $\alpha$ is given by 1.52 (for $T=0.4$). Using
this value and Eq.~(\ref{eqa4}) we can calculate $F_s(q,t)$
and compare it to the time dependence of the self-intermediate
scattering function as determined directly from the simulation of
the system {\it in equilibrium}, i.e. for large $t_w$. This is done in
Fig.~\ref{Fig_beta_rmsT0.4}. We see that the correlator as determined from
Eq.~(\ref{eqa4}) (filled circles) describes indeed very well the data from
the simulation for long waiting times, i.e. the equilibrium
dynamics (open squares). Both functions can be fitted very well by a
stretched exponential with a stretching parameter $\beta=0.84$ (solid
line). Thus we can conclude that the stretching of $F_s(q,t_w,t_w+\tau)$
at long waiting times is related to the motion of the individual clusters
that have a size distribution given by an exponential, and therefore should not be related to any kind of glassy dynamics.

\vspace*{4mm}
{\bf Acknowledgments:}
We thank R. Jullien, H-S. Ma 
for providing the code for the DLCA-DEF and
E. Del Gado and L. Cipelletti for useful discussions. This work has been
supported by the French Ministry of Education through  ANR TSANET, ANR
JCJC-CHEF.

\bibliographystyle{unsrt}

\end{document}